\documentclass{elsarticle}
\newcommand{\ACAO}{\mathrm{ACA}_0}
\newcommand{\ACA}{\mathrm{ACA}}
\newcommand{\ACAOp}{\mathrm{ACA}_0^+}

\newcommand{\LTTW}{\mathrm{LTT}_\mathrm{W}}
\newcommand{\LTTO}{\mathrm{LTT}_0}
\newcommand{\LTTOs}{\LTTO^*}

\newcommand{\TwU}{\mathrm{T}_\omega U}
\newcommand{\Tw}{\mathrm{T}_\omega}
\newcommand{\Tt}{\mathrm{T}_2}

\renewcommand{\L}{\mathrm{L}_2}
\newcommand{\s}{\mathrm{s}}
\renewcommand{\S}{\mathrm{S}}
\newcommand{\Set}[1]{\mathrm{Set} \left( {#1} \right)}

\newcommand{\brackets}[1]{\left[ \! \left[ {#1} \right] \! \right]}
\newcommand{\angles}[1]{\left\langle \! \left| {#1} \right| \! \right\rangle}
\newcommand{\sangles}[1]{\left| {#1} \right|}
\newcommand{\round}[1]{( \! | {#1} | \! )}
\newcommand{\corners}[1]{\, \ulcorner {#1} \urcorner}

\newcommand{\N}{\mathbb{N}}
\renewcommand{\s}{\operatorname{s}}
\newcommand{\EN}{\mathrm{E}_\N}

\newcommand{\IndN}{\mathrm{Ind}_\N}
\newcommand{\etaX}{\mathrm{eta}_\times}
\newcommand{\etaA}{\mathrm{eta}_\rightarrow}

\newcommand{\eqN}{=_\N}
\newcommand{\SetN}{\Set{\N}}
\newcommand{\R}{\mathrm{R}}

\newcommand{\vald}{\mathrm{valid}}
\newcommand{\type}{\ \mathrm{type}}
\newcommand{\p}{\ \mathrm{prop}}
\newcommand{\prop}{\ \mathrm{Prop}}

\newcommand{\dom}{\operatorname{dom}}

\newcommand{\Seq}{\mathbf{Seq}}

\newcommand{\vdashu}{\vdash^+}
\newcommand{\vdashw}{\vdash^-}
\newcommand{\code}{\mathrm{code}}
\newcommand{\decode}{\mathrm{decode}}
\newcommand{\PP}{\mathbf{P}}
\newcommand{\QQ}{\mathbf{Q}_1}
\newcommand{\RR}{\mathbf{Q}_2}
\newcommand{\TT}{\mathscr{T}}

\newcommand{\plus}{\ \mathtt{plus} \ }
\newcommand{\ttimes}{\ \mathtt{times} \ }
\newcommand{\FV}[1]{\mathrm{FV} \left( {#1} \right)}

\usepackage{amsmath}
\usepackage{amssymb}
\usepackage{mathrsfs}
\usepackage{diagrams}
\usepackage{prooftree}
\usepackage{eqnarray}
\usepackage{QED}

\newtheorem{thm}{Theorem}[section]
\newtheorem{lm}[thm]{Lemma}
\newtheorem{cor}{Corollary}[thm]
\newdefinition{df}[thm]{Definition}
\newproof{pf}{Proof}

\begin{document}

\title{Classical Predicative Logic-Enriched Type Theories\tnoteref{abbs}}

\tnotetext[abbs]{This research was supported by the UK EPSRC research grant EP/D066638/1 and research grant
F/07-537/AA of the Leverhulme Trust in the UK.}

\author{Robin Adams\corref{cor}}
\ead{robin@cs.rhul.ac.uk}
\author{Zhaohui Luo}
\ead{zhaohui@cs.rhul.ac.uk}

\address{Royal Holloway, University of London}

\cortext[cor]{Corresponding author.  Address: Department of Computer Science, Royal Holloway, University of London, Egham Hill, Egham, Surrey, TW20 0EX, United Kingdom.  Tel: +44 1784 443421.  Fax: +44 1784 439786}

\begin{abstract}
A logic-enriched type theory (LTT) is a type theory extended with a primitive mechanism for forming and proving propositions.  We construct two LTTs, named $\LTTO$ and $\LTTOs$, which we claim correspond closely to the classical predicative systems of second order arithmetic $\ACAO$ and $\ACA$.  We justify this claim by translating each second-order system into the corresponding LTT, and proving that these translations are conservative.
This is part of an ongoing research project to investigate how LTTs may be used to formalise different approaches to the foundations of mathematics.

The two LTTs we construct are subsystems of the logic-enriched type theory $\LTTW$, which is intended to formalise the classical predicative foundation presented by Herman Weyl in his monograph \emph{Das Kontinuum}.  The system $\ACAO$ has also been claimed to correspond to Weyl's foundation.  By casting $\ACAO$ and $\ACA$ as LTTs, we are able to compare them with $\LTTW$.  It is a consequence of the work in this paper that $\LTTW$ is strictly stronger than $\ACAO$.

The conservativity proof makes use of a novel technique for proving one LTT conservative over another, involving defining an interpretation of the stronger system out of the expressions of the weaker.  This technique should be applicable in a wide variety of different cases outside the present work.
\end{abstract}


\maketitle

\section{Introduction}
A lot of research in the field of mathematical logic has been devoted to constructing formal theories intended to capture various schools of thought in the foundations of mathematics.  In particular, the project of Reverse Mathematics \cite{simpson:sosoa} has provided an extremely detailed analysis of many theories in the language of \emph{second-order arithmetic} $\L$.  It has been argued that the theories studied correspond closely to different foundational schools; in particular, that the  classical, predicative foundation presented by Hermann Weyl in his monograph \emph{Das Kontinuum} \cite{weyl:kontinuum} is captured by the theory $\ACAO$ \cite{feferman:kontinuum}.

The systems of logic known as \emph{dependent type theories} have also received a lot of attention, and in particular have proven to offer many practical benefits when
used as the basis of the computer systems known as \emph{proof checkers} or \emph{proof assistants}.  Type theories divide the world of mathematical objects
into \emph{types}.  They offer much more expressive power than second-order arithmetic: we are able to speak, not just of natural numbers and sets of natural
numbers, but also about (e.g.) sets of sets, lists, trees, and functions from any of these types to any of them.  However, so far, type theories have been used
almost exclusively to represent \emph{constructive} mathematics.

More recently, the concept of a \emph{logic-enriched type theory} has been developed.  A logic-enriched type theory is a type theory augmented with a separate, primitive mechanism for forming and proving propositions.  It thus has two components or `worlds': a type-theoretic component, consisting of objects collected into types, and a logical component, for reasoning about these objects.  LTTs have been used to investigate the relationships between type theories and set theories \cite{ag:cpdtt,ga:gticst}, and by the present authors \cite{al:wpcm2, al:wpcm} to formalise the predicative foundation for mathematics presented by Hermann Weyl in \emph{Das Kontinuum} \cite{weyl:kontinuum}.

There is reason to believe that LTTs may offer some of the advantages of both traditional logical systems, and type theories.  They share with type theories the rich type structure and inbuilt notion of computation that have proven to be of great benefit for formalisation in practice.  At the same time, they offer the flexibility in choice of axioms that we are used to in traditional logical systems: it is possible, for example, to add excluded middle to the logical component without changing the type-theoretic component.

This paper is part of an ongoing research project to construct a hierarchy of LTTs, similar to the hierarchy of second-order systems in Reverse Mathematics. We hope thereby to investigate how LTTs may be used to represent different schools of thought in the foundations of mathematics, and to understand the effect that changes in the design of an LTT have on its set of definable objects and provable theorems.

In this paper, we construct two LTTs that capture two second-order systems that are closely related to the foundation of \emph{Das Kontinuum}: $\ACAO$ and $\ACA$.  We construct two LTTs, which we name $\LTTO$ and $\LTTOs$.  These are more expressive than a second-order system: the type-theoretic component of each features types of natural numbers, pairs, functions of all orders, and sets of all orders.

Our aim in this paper is to show that adding this expressive power is `safe'; that is, that we have not thereby increased the proof-theoretic strength of the
system.  We do this as follows.  Let us say that a proposition of $\LTTO$ is \emph{second-order} iff it uses no types other than $\N$ (the type of natural
numbers) and $\Set{\N}$ (the type of sets of natural numbers).  We define a translation from $\ACAO$ onto the second-order propositions of $\LTTO$, and prove
that the translation is \emph{conservative}; that is, 
a formula of $\L$ is provable in $\ACAO$ if and only if its translation is provable in $\LTTO$.

The current authors have previously \cite{al:wpcm2,al:wpcm} presented a new system intended to capture Weyl's foundation, which we named $\LTTW$.  We argued
there that $\LTTW$ captures Weyl's foundation very closely, and described how all the definitions and results in \emph{Das Kontinuum} have been formalised in
$\LTTW$ using a proof assistant.  The two LTTs that we construct in this paper are both subsystems of $\LTTW$.  As a consequence of the work in this paper, we
now know that $\LTTW$ is strictly stronger than $\ACAO$, and at least as strong as $\ACA$.

We argue that, compared with $\ACAO$ and $\ACA$, $\LTTW$ corresponds more closely to the system presented in \emph{Das Kontinuum}.  This is not a claim that can be proven formally, as there is no formal definition of Weyl's foundation, but we can advance evidence for it.  In our previous paper, we pointed out the extreme similarity between the presentation in \emph{Das Kontinuum} and the definition of $\LTTW$, and described one construction in \emph{Das Kontinuum} --- the construction of $K(n) = \{ X \mid X \mbox{ has at least } n \mbox{ elements} \}$ --- that cannot be done `as directly' in any of the second order systems.  Here, we strengthen the justification for this claim: we show that $K$ is expressed by a term in $\LTTW$ that cannot be formed in either $\LTTO$ or $\LTTOs$.

The majority of this paper is taken up with proving the conservativity results.  
Our method for proving the conservativity of $\LTTO$ over $\ACAO$ is as follows.  We first define a subsystem $\Tt$ of $\LTTO$ which has just two types, $\N$
and $\Set{\N}$.  and show that $\LTTO$ is conservative over $\Tt$.

We then construct infinitely many subsystems of $\LTTO$ between $\Tt$ and $\LTTO$.  We prove that, for each of these subsystems $S$ and $T$, whenever $S$ is a
subsystem of $T$, then $T$ is conservative over $S$.  We do this by defining an interpretation of the judgements of $T$ in terms of the expressions of $S$. 
Informally, we can think of this as giving a way of reading the judgements of $T$ as \emph{statements about $S$}.  We show that this interpretation satisfies
two properties:
\begin{itemize}
\item
Every derivable judgement of $T$ is true.
\item
Every judgement of $S$ that is true is derivable in $S$.
\end{itemize}
It follows that, if a judgement of $S$ is derivable in $T$, then it is derivable in $S$.

The proof thus makes use of an original technique which should be of interest in its own right, and which we expect to be applicable in a wide variety of contexts for proving one LTT or type theory conservative over another.  In particular, we shall show how it can be adapted to provide a direct proof that $\ACAO$ is conservative over Peano Arithmetic.

\subsection{Outline}
In Section 2 of this paper, we describe the subsystems of second order arithmetic that we shall consider, and compare them informally with Weyl's system.  In Section 3, we give the formal definition of $\LTTW$ and its two subsystems, and define the translation from second order arithmetic into the LTTs.  In Section 4, we prove that this translation is conservative in the case of $\ACAO$ and $\Tt$.  In Section 5, we prove that $\LTTO$ is conservative over $\Tt$.  Finally, in Section 6, we indicate how the proof can be modified to prove the conservativity of $\LTTOs$ over $\ACA$, and discuss the possibility of constructing a subsystem of $\LTTW$ conservative over $\ACAOp$, and the conservativity of $\ACAO$ over Peano Arithmetic.

\paragraph{Notation}
We shall stick to the following convention throughout this paper.  Capital letters from the beginning of the Latin alphabet ($A$, $B$, $C$, \ldots) shall denote types.  Capital letters from the middle ($K$, $L$, $M$, $N$, \ldots) shall denote terms.  Capital letters from just after the middle ($P$, $Q$) shall denote names of small propositions.  Lower-case letters ($x$, $y$, $z$, \ldots) shall denote variables, except $t$, which we reserve for terms of the language of second-order arithmetic.  Lower-case letters from the middle of the Greek alphabet ($\phi$, $\psi$, $\chi$, \ldots) shall denote propositions.

We shall be dealing with partial functions throughout this paper.  We write $X \simeq Y$ to denote that the expression $X$ is defined if and only if $Y$ is defined, in which case they are equal.  Given a function $v$, we write $v[x:=a]$ for the function $v'$ with domain $\dom v \cup \{x\}$, such that $v'(x) = a$, and $v'(y) = v(y)$ for $y \neq x$.
We write $\FV{X}$ for the set of free variables in the expression $X$.

\section{Background}

\subsection{Weyl's \emph{Das Kontinuum}}
In 1918, Herman Weyl wrote the monograph \emph{Das Kontinuum} \cite{weyl:kontinuum}, which presented a semi-formal system intended to provide a predicative foundation for
mathematics.  Weyl's system consists of a set of `principles' by which sets, functions and propositions may be introduced.  In particular, if we have formed the proposition $\phi$, we may introduce the set $\{ x \mid \phi \}$, provided that $\phi$ does not involve any quantification over sets.  Impredicative definitions are thus impossible in Weyl's system.  His concern was to show how much of mathematics --- in particular, how much of analysis --- could still be retained under such a restriction.

At the time of writing \emph{Das Kontinuum} in 1918, Weyl agreed with Whitehead and Russell's opinion \cite{wr:pm} that the source of the famous paradoxes in set theory was
the presence of \emph{impredicative definitions} --- definitions that involved a certain kind of \emph{vicious circle}.  In particular, when we introduce a set $R$ with the definition
\begin{equation}
\label{eq:dfset}
 R = \{ x \mid \phi \}
\end{equation}
then the definition is impredicative if either $x$ or any of the bound variables in $\phi$ ranges over a collection that includes the set $R$ itself.

In Weyl's foundation, mathematical objects are divided into \emph{categories}.  A category can be \emph{basic} or \emph{ideal}.  Given any category $A$, there is the ideal category $\Set{A}$\footnote{The notation here is ours, not Weyl's.} of \emph{sets} whose members are objects of category $A$.  In a definition of the form (\ref{eq:dfset}), we may only quantify over \emph{basic} categories.  In particular, we may not quantify over any category of the form $\Set{A}$.  It is in this manner that impredicative definitions are excluded.

If we bar impredicative definitions, we are unable to define many objects, such as the least upper bound of a bounded set of reals.  We must thus either find an
alternative way to introduce these objects, or do without them.  
Russell and Whitehead chose the former course, with their Axiom of Reducibility.  The monograph \emph{Das Kontinuum} was Weyl's attempt to follow the latter
course: to show how much of classical mathematics could be preserved while excluding impredicative definitions.

\subsection{Subsystems of Second Order Arithmetic}

We are concerned in this paper with two subsystems of second order arithmetic, $\ACAO$ and $\ACA$.
The letters ACA stand for `arithmetical comprehension axiom'.  The system $\ACAO$ is investigated in great detail in Simpson \cite{simpson:sosoa}.  These two systems are theories in the language of second order arithmetic, a language for describing natural numbers and sets of natural numbers.  We now introduce this language formally.

\begin{df}[Language of Second Order Arithmetic]
The \emph{language \\ of second order arithmetic} $\L$ is defined as follows.

There are two countably infinite, disjoint sets of \emph{variables}: the \emph{number variables} $x$, $y$, $z$, \ldots, intended to range over natural numbers; and the \emph{set variables} $X$, $Y$, $Z$, \ldots, intended to range over sets of natural numbers.

The \emph{terms} and \emph{propositions} of second order arithmetic are given by the following grammar:
\[ \begin{array}{lrcl}
    \mbox{Term} & t & ::= & x \mid 0 \mid \S t \mid t + t \mid t \cdot t \\
\mbox{Proposition} & \phi & ::= & t = t \mid t \in X \mid \bot \mid \phi \supset \phi \mid \forall x \phi \mid \forall X \phi
   \end{array} \]
We define $\neg$, $\wedge$, $\vee$, $\leftrightarrow$ and $\exists$ in terms of $\bot$, $\supset$ and $\forall$ as usual.

A proposition is \emph{arithmetic} iff no set quantifier $\forall X$ occurs within it.
\end{df}

\subsubsection{$\ACAO$}
The system $\ACAO$ has been very well studied.  In particular, it has played a major role in the
project of Reverse Mathematics \cite{simpson:sosoa}.  It has often been argued that $\ACAO$ is closely related to Weyl's foundation; for example, Feferman \cite{feferman:kontinuum} calls it `a modern formulation of Weyl's system', and Brown and Simpson \cite{bs:hb} write `$\ACAO$ isolates the same portion of mathematical practice which was identified as `predicative analysis' by Herman Weyl in his famous monograph Das Kontinuum'.

It is known that $\ACAO$ is conservative over Peano Arithmetic (PA); a model-theoretic proof is given in Simpson \cite{simpson:sosoa}, and a proof-theoretic proof can be given along the lines of Shoenfield \cite{shoenfield:rcp}.  A novel proof of this result shall be given in Section \ref{section:ACAOPA}.

The axioms of $\ACAO$ are as follows:
\begin{itemize}
 \item The \emph{Peano axioms} --- the axioms of Peano Arithmetic, minus the induction axioms:
\begin{gather*}
 \S x \neq 0 \\
\S x = \S y \supset x = y \\
x + 0 = x \\
x + \S y = \S (x + y) \\
x \cdot 0 = 0 \\
x \cdot \S y = x \cdot y + x
\addtocounter{equation}{1}
\end{gather*}
\item
The \emph{arithmetical comprehension axiom schema}: for every arithmetic \\ proposition $\phi$ in which $X$ does not occur free,
$\exists X \forall x (x \in X \leftrightarrow \phi)$.
\item
The \emph{set induction axiom}:
$0 \in X \supset \forall x (x \in X \supset \S x \in X) \supset \forall x. x \in X$.
\end{itemize}

\subsubsection{$\ACA$}
The system $\ACA$ is formed by extending $\ACAO$ with the full \emph{induction axiom schema}: for every proposition $\phi$,
\[ [0/x]\phi \supset \forall x (\phi \supset [\S x / x] \phi) \supset \forall x \phi \enspace . \]
An argument could be made for $\ACA$ being a better representation of the foundation in \emph{Das Kontinuum} than $\ACAO$, because --- as we shall argue in Section \ref{section:weylssystem} --- Weyl makes use of an induction principle
that is stronger than that of $\ACAO$.

The system $\ACA$ has not been studied in the literature as much as $\ACAO$.
A few facts about $\ACA$ are known: its proof-theoretic ordinal is $\varepsilon_{\varepsilon_0}$, and it can prove the consistency of $\ACAO$.  See \cite{as:mtoapt} for the proof of these results and an analysis of the set of models of $\ACA$.

\subsection{\emph{Das Kontinuum} and Subsystems of Second Order Arithmetic Compared}
\label{section:weylssystem}
There has been quite some argument over how well Weyl's foundation is captured by a subsystem of second order arithmetic.  Feferman \cite{feferman:wv} has argued strongly in favour of $\ACAO$, or a system very like it, being a modern formulation of Weyl's system.  

This argument cannot be settled formally, as Weyl did not give a formal definition of his system.  However, in the authors' view, Weyl's system exceeds both $\ACAO$ and $\ACA$, for the following reasons:
\begin{enumerate}
 \item Weyl intended his system to be more than second order.  He allowed the category $\Set{B}$ to be formed for \emph{any} category $B$, basic or ideal.  Thus, for example, we can form the categories $\Set{\Set{\N}}$, $\Set{\Set{\Set{\N}}}$, and so forth.
\item Weyl intended the principle of induction to apply to all propositions, arithmetic or not.
\end{enumerate}
We justify this by showing a place where Weyl explicitly defines a function of category $\Set{\Set{A}} \rightarrow \Set{\Set{A}}$, and three places where he proves a non-arithmetic proposition by induction.

The former occurs \cite[p.39]{weyl:continuum} with the definition of the cardinality of a set.  Weyl defines a function $d:\Set{\Set{A}} \rightarrow \Set{\Set{A}}$ by
\[ d(\TT) = \{ X \mid \exists x \in X. X \setminus \{x\} \in \TT \} \enspace . \]
This function is then iterated, to form the function \\ $d^n(\TT) = \{ X \mid n \mbox{ elements may be removed from } X \mbox{ to form an element of } \TT \}$.  Weyl goes on to argue that $d^n(\mathscr{U})$ denotes the set of all sets with at least $n$ elements (where $\mathscr{U}$ is the set of all subsets of $A$).  He defines the proposition $a(n,X)$, `$X$ has at least $n$ elements', by
\[ a(n,X) \equiv X \in d^n(\mathscr{U}) \]

Various results about this definition are later proved \cite[p.55]{weyl:continuum}, such as:
\begin{quote}
 If $X$ has at least $n+1$ elements, then $X$ has at least $n$ elements.
\end{quote}
This is not an arithmetic proposition (it involves quantification over $X$), but it is proven by induction on $n$.

Similarly, the non-arithmetic proposition `If $X$ is a subset of $E$ and $X$ consists of at least $n$ elements, then $E$ also consists of at least $n$ elements'
\cite[p.56]{weyl:continuum} is proven by induction, as is the lemma concerning \emph{substitution of elements} \cite[p.56]{weyl:continuum}: `If a new object
[\ldots] is substituted for one of the elements of a set $X$ which consists of at least $n$ elements [\ldots], then the modified set $X^*$ also consists of at
least $n$ elements.'

Thus, Weyl's method of defining $a(n,X)$ involves third-order sets; the application of the Principle of Iteration to third-order sets; and proof by
induction of a proposition that quantifies over sets.  These are all expressed by primitive constructs in $\LTTW$, but not in $\LTTO$ or $\LTTOs$ (we discuss
this point further in Section \ref{section:ltts}).  

When we have proven the conservativity of $\LTTO$ and $\LTTOs$ over $\ACAO$ and $\ACA$ respectively, we will have justified our claim that
Weyl's system is stronger than $\ACAO$; and, if our conjecture that $\LTTW$ is stronger than $\LTTOs$ is correct, that Weyl's system is
stronger than $\ACA$.

\section{Logic-Enriched Type Theories}

In this section, we introduce the logic-enriched type theory $\LTTW$ and the two subsystems with which we are concerned.

Logic-enriched type theories (LTTs) were introduced by Aczel and Gambino \cite{ag:cpdtt,ga:gticst} to study the relationship between type theories and set theories.  An LTT is a formal system consisting of two parts: the \emph{type-theory component}, which deals with \emph{terms} and \emph{types}; and the \emph{logical component}, which deals with \emph{propositions}.

\subsection{$\LTTW$}

The system $\LTTW$ is a logic-enriched type theory designed to represent the mathematical foundation given in \emph{Das Kontinuum}.  It was introduced in Adams and Luo \cite{al:wpcm,al:wpcm2}.

\subsubsection{Type-Theoretic Component}
Its type-theoretic component has the following types.
\begin{itemize}
 \item There is a type $\mathbb{N}$ of natural numbers.  0 is a natural number; and, for any natural number $N$, the \emph{successor} of $N$, $\s N$, is a natural number.
\item For any types $A$ and $B$, we may form the type $A \times B$.  Its terms are pairs $(M,N)_{A \times B}$ consisting of a term $M$ of $A$ and a term $N$ of $B$.  For any term $M : A \times B$, we can construct the term $\pi_1^{A \times B}(M)$ denoting its first component, and the term $\pi_2^{A \times B}(M)$ denoting its second component.
\item For any types $A$ and $B$, we may form the type $A \rightarrow B$ of functions from $A$ to $B$.  Its terms have the form $\lambda x:A.M:B$, denoting the function which, given $N : A$, returns the term $[N/x]M : B$.  Given $M : A \rightarrow B$ and $N : A$, we may construct the term $M(N)_{A \rightarrow B}$ to denote the value of the function $M$ when applied to $N$.
\item For any type $A$, we may form the type $\Set{A}$ of \emph{sets} of terms of $A$.  Its terms have the form $\{ x : A \mid P \}$, where $P$ is a name of a small proposition, denoting the set of all $M:A$ for which the proposition named by $[M/x]P$ is true.
\end{itemize}

We divide the types into \emph{small} and \emph{large} types, reflecting Weyl's division of categories into \emph{basic} and \emph{ideal} categories.  When we introduce a set $\{ x : A \mid P \}$, the proposition $P$ may quantify over the small types, but not over the large types.  The small types are defined inductively by:
\begin{itemize}
 \item $\N$ is a small type.
\item If $A$ and $B$ are small types, then $A \times B$ is a small type.
\end{itemize}

We effect this division by introducing a \emph{type universe} $U$, whose terms are \emph{names} of the small types.  There is a term $\hat{\N} : U$ which is the name of $\N$; and, if $M : U$ names $A$ and $N : U$ names $B$, then there is a term $M \hat{\times} N : U$ that names $A \times B$.  We write $T(M)$ for the type named by $M$.

We can also \emph{eliminate} $\N$ over any family of types; that is, if $A[x]$ is a type depending on $x : \N$, we can define by recursion a function $f$ such that $f(x) : A[x]$ for all $x : \N$.  The term
\[ \EN([x]A, L, [x,y]M, N) \]
is intended to denote the value $f(N)$, where $f$ is the function defined by recursion thus:
\begin{eqnarray*}
 f(n) & : & [n/x]A \qquad \mbox{for all } n : \N \\
f(0) & = & L \\
f(n+1) & = & [n/x, f(n)/y]M
\addtocounter{equation}{1} \end{eqnarray*}

\paragraph{Remark}
We choose to label the terms 
\[ (M,N)_{A \times B}, \pi_1^{A \times B}(M), \pi_2^{A \times B}(M), \lambda x:A.M:B \mbox{ and } M(N)_{A \times B} \]
with the types $A$ and $B$.  This is for technical reasons only; it makes the interpretations we introduce in Section \ref{section:conservativity} easier to define.  We shall often omit these labels when writing terms.  We shall also often write $MN$ for $M(N)$.

\subsubsection{Logical Component}
The logical component of $\LTTW$ contains propositions built up as follows:
\begin{itemize}
 \item If $M$ and $N$ are objects of the small type $T(L)$, then $M =_L N$ is a proposition.
\item $\bot$ is a proposition.
\item If $\phi$ and $\psi$ are propositions, then $\phi \supset \psi$ is a proposition.
\item If $A$ is a type and $\phi$ a proposition, then $\forall x:A.\phi$ is a proposition.
\end{itemize}
We define the other logical connectives as follows:
\begin{eqnarray*}
 \neg \phi & \equiv & \phi \supset \bot \\
\phi \wedge \psi & \equiv & \neg (\phi \supset \neg \psi) \\
\phi \vee \psi & \equiv & \neg \phi \supset \psi \\
\phi \leftrightarrow \psi & \equiv & (\phi \supset \psi) \wedge (\psi \supset \phi) \\
\exists x:A.\phi & \equiv & \neg \forall x:A. \neg \phi
\addtocounter{equation}{1} \end{eqnarray*}

We call a proposition $\phi$ \emph{small} iff, for every quantifier $\forall x:A$ that occurs in $\phi$, the type $A$ is a small type.
We wish it to be the case that, when we introduce a set of type $\Set{A}$, the proposition we use to do so must be a \emph{small} proposition.  

We achieve this by introducing a \emph{propositional universe} `prop', which will be the collection of names of the small propositions.  We shall introduce
a new judgement form $\Gamma \vdash P \p$, denoting that $P$ is the name of a small proposition, and rules that guarantee:
\begin{itemize}
 \item If $M$ and $N$ are objects of the small type $T(L)$, then $M \hat{=}_L N$ is the name of $M =_L N$.
\item $\hat{\bot}$ is the name of $\bot$.
\item If $P$ names $\phi$ and $Q$ names $\psi$, then $P \hat{\supset} Q$ is the name of $\phi \supset \psi$.
\item If $M:U$ names the small type $A$ and $P$ names $\phi$, then $\hat{\forall} x : M. P$ names $\forall x:A.\phi$.
\end{itemize}
We denote by $V(P)$ the small proposition named by $P$.  We shall, in the sequel, often write just `small proposition' when we should strictly write `name of small proposition'.

We use `expression' to mean a type, term, small proposition or proposition.
We identify expressions up to $\alpha$-conversion.  We denote by $[M/x]X$ the result of substituting the term $M$ for the variable $x$ in the expression $X$, avoiding variable capture.

\subsubsection{Judgements and Rules of Deduction}

A \emph{context} in $\LTTW$ has the form $x_1 : A_1, \ldots, x_n : A_n$, where the $x_i$s are distinct variables and each $A_i$ is a type.
There are ten judgement forms in $\LTTW$:
\begin{itemize}
 \item $\Gamma \vdash \vald$, denoting that $\Gamma$ is a valid context.
\item $\Gamma \vdash A \type$, denoting that $A$ is a well-formed type under the context $\Gamma$.
\item $\Gamma \vdash A = B$, denoting that $A$ and $B$ are equal types.
\item $\Gamma \vdash M : A$, denoting that $M$ is a term of type $A$.
\item $\Gamma \vdash M = N : A$, denoting that $M$ and $N$ are equal terms of type $A$.
\item $\Gamma \vdash P \p$, denoting that $P$ is a well-formed name of a small proposition.
\item $\Gamma \vdash P = Q$, denoting that $P$ and $Q$ are equal names of small propositions.
\item $\Gamma \vdash \phi \prop$, denoting that $\phi$ is a well-formed proposition.
\item $\Gamma \vdash \phi = \psi$, denoting that $\phi$ and $\psi$ are equal propositions.
\item $\Gamma \vdash \phi_1, \ldots, \phi_n \Rightarrow \psi$, denoting that the propositions $\phi_1$, \ldots, $\phi_n$ entail the proposition $\psi$.
\end{itemize}

The rules of deduction of $\LTTW$ are given in full in Appendix \ref{app:lttw}.
They consist of the introduction, elimination and computation rules for the types of $\LTTW$, the rules for classical predicate logic, and the following rule for performing \emph{induction} over $\N$:

\[ (\IndN) \begin{prooftree}
\begin{array}{cc}
            \Gamma, x : \N \vdash \phi \prop
& \Gamma \vdash N : \N \\
\Gamma \vdash \Phi \Rightarrow [0/x]\phi
& \Gamma, x : \N \vdash \Phi, \phi \Rightarrow [\s x/x]\phi
\end{array}
\justifies
\Gamma \vdash \Phi \Rightarrow [N/x]\phi
           \end{prooftree} \]


\subsection{$\LTTO$ and $\LTTOs$}
\label{section:ltts}
We now construct two subsystems of $\LTTW$, which we shall call $\LTTO$ and $\LTTOs$, that correspond to $\ACAO$ and $\ACA$ respectively.  These
subsystems are formed by changing:
\begin{itemize}
 \item the class of types over which $\N$ may be eliminated (that is, the class of types $A$ that may occur in $\EN([x]A, L, [x,y]M, N)$;
\item the class of propositions that may be proved by induction (that is, the class of propositions $\phi$ that may occur in an instance of $(\IndN)$).
\end{itemize}
In $\LTTW$, we may eliminate $\N$ over any type, and any proposition may be proved by induction.  We form our three subsystems by weakening these two classes,
as shown in Table \ref{table:LTTs}

\begin{table}
\begin{tabular}{ccc}
& Types over which & Propositions provable by induction \\
& $\N$ may be eliminated & \\
\hline
$\LTTW$ & all & all \\
\hline
$\LTTO$ & small types & small propositions \\
\hline
$\LTTOs$ & small types & propositions involving quantification \\
& & over small types and $\Set{N}$
\end{tabular}
\caption{Subsystems of $\LTTW$}
\label{table:LTTs}
\end{table}

This is achieved as follows.
\begin{enumerate}
\item We construct $\LTTO$ by modifying $\LTTW$ as follows.
\begin{itemize}
\item Whenever a term $\EN([x]A, L, [x,y]M, N)$ is formed, then $A$ must have the form $T(K)$.
\item 
Whenever an instance of the rule $(\IndN)$ is used, the proposition $\phi$ must have the form $V(P)$.
\item Whenever an instance of the rule (subst), $(\etaX)$ or $(\etaA)$ is used, the proposition $\phi$ must not contain a quantifier $\forall x:A$ over any type $A$ that contains the symbol $U$.
\item We also add as an axiom that $\S M \neq 0$ for $M : \N$.
\end{itemize}
\item Let us say that a proposition $\phi$ is \emph{analytic} iff, for every quantifier $\forall x:A$ in $\phi$, $A$ either has the form $T(M)$ or $A \equiv \Set{\N}$.  We construct $\LTTOs$ from $\LTTO$ by allowing $(\IndN)$ to be used whenever $\phi$ is an analytic proposition.
\end{enumerate}
The formal definitions of both these systems are given in Appendices \ref{app:ltto} and \ref{app:lttos}.

\paragraph{Remarks}
\begin{enumerate}
\item
Peano's fourth axiom, that $\S M \neq_{\hat{\N}} 0$ for any $M : \N$, is provable in $\LTTW$; see \cite{al:wpcm2} for a proof.  It is not provable in $\LTTO$ or
$\LTTOs$.  This can be shown by a similar method to Smith \cite{smith:ipfamltt} by constructing a model of $\LTTOs$ in which every small type is interpreted by a set that has exactly
one element.
\item
We can now justify further our claim in Section \ref{section:weylssystem} that Weyl's definition of $a(n,X)$ 
uses the primitive concepts of $\LTTW$ that are not present in either $\LTTO$ or $\LTTOs$.

The definitions of $d$ and $a$ are straightforward to formalise in $\LTTW$.  Given $M : U$, we have
\begin{eqnarray*}
 d_M & \equiv & \lambda \TT : \Set{\Set{T(M)}} . \\
& & \quad \{ X : \Set{T(M)} \mid \hat{\exists} x : M. (x \hat{\in} X \wedge X \setminus \{x\} \hat{\in} \TT) \} \\
a_M & \equiv & \lambda n : \N. \lambda X : \Set{T(M)} . \\
& & \quad X \in \EN([x]\Set{\Set{T(M)}}, \mathscr{U}, [x,Y] d_M(Y), n)
\addtocounter{equation}{1} \end{eqnarray*}
This is not a term in either of the subsystems of $\LTTW$, as it involves applying $\EN$ to the type $\Set{\Set{T(M)}}$.
\item
The universe $U$ contains only the types that can be built up from $\N$ and $\times$.  Its inclusion in $\LTTO$ or $\LTTOs$ therefore does not increase the
proof-theoretic strength of the system (this will be proven in Section \ref{section:TwUTw}).  This is a rare situation; in general, the inclusion of a universe
raises the strength of a type theory considerably (see for example \cite{feferman:iifpt}).  We conjecture that, if we closed $U$ under $\rightarrow$ or
$\Set{\,}$ in $\LTTO$ or $\LTTOs$, the resulting system would not be conservative over $\ACAO$ or $\ACA$ respectively.
\item
In Aczel and Gambino's original formulation of LTTs \cite{ag:cpdtt,ga:gticst}, the logical component of an LTT could depend on the type theoretic component, but not \emph{vice versa}.  We have broken that restriction with the inclusion of \emph{typed sets}: a canonical object of $\Set{A}$ has the form $\{ x : A \mid P \}$ and thus depends on a small proposition $P$.
\end{enumerate}

\subsection{Embedding Second Order Systems in Logic-Enriched Type Theories}
\label{section:embed}

There is a translation that can naturally be defined from the language of second order arithmetic $\L$ into $\LTTO$.  We map the terms of $\L$ to terms of type $\N$, first order quantifiers to quantifiers over $\N$, and second order quantifiers to quantifiers over $\Set{\N}$.
\begin{df}
\label{df:ACAOTt}
We define
\begin{itemize}
\item
for every term $t$ of $\L$, a term $\angles{t}$ of $\LTTW$;
\item
for every arithmetic formula $\phi$ of $\L$, a small proposition $\sangles{\phi}$ of $\LTTW$;
\item
for every formula $\phi$ of $\L$, a proposition $\angles{\phi}$ of $\LTTW$.
\end{itemize}
\[ 
\begin{array}{rcl}
\angles{x_i} & \equiv & x_i \\
\angles{0} & \equiv & 0 \\
\angles{t'} & \equiv & \s \angles{t} \\
\angles{s + t} & \equiv & \angles{s} \plus \angles{t} \\
\angles{s \cdot t} & \equiv & \angles{s} \ttimes \angles{t} \\
\end{array} \]
\[ \begin{array}{rcl}
\sangles{s = t} & \equiv & \angles{s} \hat{=}_{\hat{\N}} \angles{t} \\
\sangles{t \in X_i} & \equiv & \angles{t} \hat{\in}_{\hat{\N}} X_i \\
\sangles{\neg \phi} & \equiv & \hat{\neg} \sangles{\phi} \\
\sangles{\phi \supset \psi} & \equiv & \sangles{\phi} \hat{\supset} \sangles{\psi} \\
\sangles{\forall x \phi} & \equiv & \hat{\forall} x : \hat{\N}. \sangles{\phi} \\
\end{array} \qquad
 \begin{array}{rcl}
\angles{s = t} & \equiv & \angles{s} =_{\hat{\N}} \angles{t} \\
\angles{t \in X_i} & \equiv & \angles{t} \in_{\hat{\N}} X_i \\
\angles{\neg \phi} & \equiv & \neg \angles{\phi} \\
\angles{\phi \supset \psi} & \equiv & \angles{\phi} \supset \angles{\psi} \\
\angles{\forall x \phi} & \equiv & \forall x : \N. \angles{\phi} \\
\angles{\forall X \phi} & \equiv & \forall X : \Set{\N}. \angles{\phi}
\end{array} \]
where
\begin{eqnarray*}
M \plus N & \equiv & \EN([x]T(\hat{\N}), M, [x,y]\s y, N) \\
M \ttimes N & \equiv & \EN([x]T(\hat{\N}), 0, [x,y]y \plus M, N)
\addtocounter{equation}{1} \end{eqnarray*}
\end{df}

It is straightforward to show that this translation is sound, in the following sense:
\begin{thm}
\label{thm:anglesound}
Let $\Gamma$ be the context $x_1 : \N, \ldots, x_m : \N, X_1 : \Set{\N}, \ldots, X_n : \Set{\N}$.  Let $\FV{t} \subseteq \{x_1, \ldots, x_m\}$, and
$\FV{\phi} \subseteq \{x_1, \ldots, x_m, X_1, \ldots, X_n\}$.
\begin{enumerate}
\item $\Gamma \vdash \angles{t} : \N$ and $\Gamma \vdash \angles{\phi} \prop$.
\item If $\phi$ is arithmetic, then $\Gamma \vdash \sangles{\phi} \p$
and $\Gamma \vdash V(\sangles{\phi}) = \angles{\phi}$.
 \item If $\ACAO \vdash \phi$, then $\Gamma \vdash \Rightarrow \angles{\phi}$ in $\LTTO$.
 \label{three}
\item If $\ACA \vdash \phi$, then $\Gamma \vdash \Rightarrow \angles{\phi}$ in $\LTTOs$.
\label{four}
\end{enumerate}
\end{thm}

\begin{pf}
Parts 1 and 2 are proven straightforwardly by induction on $t$ and $\phi$.

For part 3, it is sufficient to prove the case where $\phi$ is an axiom of $\ACAO$.  The case of the Peano axioms is straightforward.

For the arithmetical comprehension axiom schema, let $\phi$ be an arithmetic formula in which $X$ does not occur free.  We have
\begin{eqnarray*}
 \Gamma & \vdash & \Rightarrow \forall x : \N (V(\sangles{\phi}) \leftrightarrow \angles{\phi}) & (using part 1) \\
\therefore \Gamma & \vdash & \Rightarrow \forall x : \N (x \in \{ x : \N \mid \sangles{\phi} \} \leftrightarrow \angles{\phi}) \\
\therefore \Gamma & \vdash & \Rightarrow \exists X : \Set{\N}. \forall x : \N (x \in X \leftrightarrow \angles{\phi})
\addtocounter{equation}{1} \end{eqnarray*}
as required.

The set induction axiom is shown to be provable using $(\IndN)$.

For part 4, it is sufficient to show that every instance of the full induction axiom schema is provable in $\LTTOs$.  This is easy to do using $(\IndN)$, as $\angles{\phi}$ is always an analytic proposition.
\end{pf}

\begin{cor}
$\LTTW$ is strictly stronger than $\ACAO$.  In fact, $\LTTW$ can prove the consistency of $\ACAO$.
\end{cor}

\begin{pf}
As $\ACAO$ is conservative over Peano Arithmetic \cite{simpson:sosoa}, its proof-theoretic ordinal is $\epsilon_0$.  The proof-theoretic ordinal of $\ACA$ is
$\epsilon_{\epsilon_0}$ \cite{as:mtoapt,schutte:pt}.  Therefore, $\ACA$ can prove the consistency of $\ACAO$; hence, so can $\LTTOs$; hence, so can $\LTTW$.
\end{pf}

Our aim in this paper is to prove the converse to Theorem \ref{thm:anglesound} parts \ref{three} and \ref{four}: that, whenever $\Gamma \vdash \Rightarrow \angles{\phi}$ in $\LTTO$ or $\LTTOs$, then $\phi$ is provable in the
corresponding subsystem of second order arithmetic.

\section{Conservativity of $\Tt$ over $\ACAO$}
\label{section:TtACAO}
We shall now define the system $\Tt$, which is a subsystem of $\LTTO$.
We can think of $\Tt$ as the second order fragment of $\LTTO$; that is, the part of $\LTTO$ that has just the two types $\N$ and $\Set{\N}$.

The translation $\angles{\,}$ given in the previous section is in fact a sound translation of $\ACAO$ into $\Tt$.  In this section, we shall prove that this translation is conservative; that is, if $\angles{\phi}$ is provable in $\Tt$, then $\phi$ is a theorem of $\ACAO$.

The syntax of $\Tt$ is given by the following
grammar
\[ \begin{array}{lrcl}
    \mbox{Type} & A & ::= & \N \mid \SetN \\
\mbox{Term} & M & ::= & x \mid 0 \mid \s M \mid \R(M,[x,x]M,M) \mid \{ x:\N \mid P \} \\
\mbox{Small Proposition} & P & ::= & M \hat{\eqN} M \mid \hat{\bot} \mid P \hat{\supset} P \mid \hat{\forall} x:\hat{\N} . P \mid M \hat{\in}_\N M \\
\mbox{Proposition} & \phi & ::= & M \eqN M \mid \bot \mid \phi \supset \phi \mid \forall x:A.\phi \mid V(P)
   \end{array} \]
The rules of deduction of $\Tt$ are:
\begin{enumerate}
 \item the structural rules for LTTs as given in Appendix \ref{section:LTTs};
\item the rules for predicate logic as given in Appendix \ref{section:predlog};
\item the rules for the propositional universe as given in Appendix \ref{section:propuni}, with the rules for universal quantification
replaced with the rules in Figure \ref{fig:tt1};
\item the rules for equality given in Appendix \ref{section:equality}, restricted to the type $\N$;
\item the rules for sets given in Appendix \ref{section:sets}, restricted to the type $\N$;
\item the rules for natural numbers given in Figure \ref{fig:tt}.
\end{enumerate}

\begin{figure}
\[ \begin{prooftree}
\Gamma, x : \mathbb{N} \vdash P \prop
\justifies
\Gamma \vdash \hat{\forall} x : \hat{\N}. P \prop
\end{prooftree}
\qquad
\begin{prooftree}
\Gamma, x : \mathbb{N} \vdash P = Q
\justifies
\Gamma \vdash (\hat{\forall} x : \hat{\N}. P) = (\hat{\forall} x . \hat{\N}. Q)
\end{prooftree} \]
\[ \begin{prooftree}
\Gamma, x : \N \vdash P \prop
\justifies
\Gamma \vdash V(\hat{\forall} x : \hat{\N}. P) = \forall x : \N. V(P)
\end{prooftree} \]
\caption{Rules of Deduction for Small Universal Quantification in $\Tt$}
\label{fig:tt1}
\end{figure}

\begin{figure}
\[ \begin{prooftree}
    \Gamma \vdash \vald
\justifies
\Gamma \vdash \N \type
   \end{prooftree} \qquad
\begin{prooftree}
\Gamma \vdash \vald
\justifies
\Gamma \vdash 0 : \N
\end{prooftree}
\qquad
\begin{prooftree}
 \Gamma \vdash M : \N
\justifies
\Gamma \vdash \s M : \N
\end{prooftree}
\qquad
\begin{prooftree}
 \Gamma \vdash M = M' : \N
\justifies
\Gamma \vdash \s M = \s M' : \N
\end{prooftree} \]

\[ \begin{prooftree}
\begin{array}{c}
    \Gamma \vdash L : \N
\\
\Gamma, x : \N, y : \N \vdash M : \N
\\
\Gamma \vdash N : \N
\end{array}
\justifies
\Gamma \vdash \R(L,[x,y]M,N) : \N
   \end{prooftree}
\qquad
\begin{prooftree}
\begin{array}{c}
 \Gamma \vdash L = L' : \N
\\
\Gamma, x : \N, y : \N \vdash M = M' : \N
\\
\Gamma \vdash N = N' : \N
\end{array}
\justifies
\Gamma \vdash \R(L,[x,y]M,N) = \R(L',[x,y]M',N') : \N
\end{prooftree} \]

\[ \begin{prooftree}
\begin{array}{c}
\Gamma \vdash L : \N
\\
\Gamma, x : \N, y : \N \vdash M : \N
\end{array}
\justifies
\Gamma \vdash \R(L,[x,y]M,0) = L : \N
   \end{prooftree}
\qquad
\begin{prooftree}
\begin{array}{c}
    \Gamma \vdash L : \N
\\
\Gamma, x : \N, y : \N \vdash M : \N
\\
\Gamma \vdash N : \N
\end{array}
\justifies
\begin{array}{l}
\Gamma \vdash \R(L,[x,y]M,\s N) \\
\quad = [N/x, \R(L, [x,y]M, N)/y]M : \N 
\end{array}
\end{prooftree} \]

\[ (\IndN) \; \begin{prooftree}
\begin{array}{cc}
\Gamma, x : \N \vdash P \prop
&
\Gamma \vdash N : \N \\
\Gamma \vdash \Phi \Rightarrow V([0/x]P)
&
\Gamma \vdash \Phi, V(P) \Rightarrow V([\s x/x]P)
\end{array}
\justifies
\Gamma \vdash \Phi \Rightarrow V([N/x]P)
\end{prooftree} \]
\caption{Rules of Deduction for Natural Numbers in $\Tt$}
\label{fig:tt}
\end{figure}

\paragraph{Note}
$\Tt$ does not contain the universe $U$.  The symbol $\hat{\N}$ therefore is not a term in $\Tt$, and cannot occur on its own,
but only as part of a small proposition $\hat{\forall} x : \hat{\N}. P$.

In $\LTTO$, we could define functions by recursion into any small type; in $\Tt$, we can only define by recursion functions from $\N$ to $\N$.  This is achieved
by the constructor $\R$.
The term $\R(L, [x,y]M, N)$ is intended to denote the value $f(N)$, where $f : \N \rightarrow \N$ is defined by recursion thus:
\begin{eqnarray*}
f(0) & = & L \\
f(n+1) & = & [n/x,f(n)/y]M
\end{eqnarray*}

The system $\Tt$ may be considered a subsystem of $\LTTO$ if we identify $\R(L,[x,y]M,N)$ with $\EN([x]\hat{\N}, L, [x,y]M, N)$;
$M \eqN N$ with $M =_{\hat{\N}} N$; and
$M \hat{\eqN} N$ with $M \hat{=}_{\hat{\N}} N$.

The translation given in Section \ref{section:embed} is a sound translation from $\ACAO$ into $\Tt$.
\begin{thm}
Let $\Gamma$ and $\phi$ be as in Theorem \ref{thm:anglesound}.  If $\ACAO \vdash \phi$, then $\Gamma \vdash \Rightarrow \angles{\phi}$ in $\Tt$.
\end{thm}

\begin{pf}
Similar to the proof of Theorem \ref{thm:anglesound}(3).
\end{pf}
We now wish to show that the converse holds.

We shall do this by defining the following translation $\Phi$ from $\Tt$ to $\ACAO$.  Let
\[ \Gamma \equiv x_1 : \N, \ldots, x_n : \N, X_1 : \SetN, \ldots, X_m : \SetN \enspace . \]
We shall define:
\begin{enumerate}
 \item whenever $\Gamma \vdash M : \N$, an arithmetic formula $t \corners{=M}$ such that
\[ \ACAO \vdash \exists ! x. x \corners{= M} \enspace . \]
The intention is that $M$ is interpreted as the unique number $x$ for which $x \corners{= M}$ is true.
\item whenever $\Gamma \vdash M : \SetN$, an arithmetic formula $t \corners{\in M}$ such that
\[ \ACAO \vdash \exists X \forall x (x \in X \leftrightarrow x \corners{\in M}) \enspace , \]
The intention is that $M$ is interpreted as the unique set $X$ whose members are the numbers $x$ such that $x \corners{\in M}$ is true.
\item for every small proposition $P$ such that $\Gamma \vdash P \p$, an arithmetic formula $\corners{P}$.
\item for every proposition $\phi$ such that $\Gamma \vdash \phi \prop$, a formula $\corners{\phi}$.
\end{enumerate}

The definition is given in Figure \ref{fig:TtACAO}.
\begin{figure}
\paragraph{Numbers}
\begin{eqnarray*}
t \corners{= x_i} & \equiv & t = x_i \\
t \corners{= 0} & \equiv & t = 0 \\
t \corners{= \s M} & \equiv & \exists x (x \corners{= M} \wedge t = \S x) \\
t \corners{= \R(L, [u,v]M, N)} & \equiv & \exists n. \exists s \in \Seq ( n \corners{= N} \wedge (n,t) \in s \\
& & \wedge \forall l ((0,l) \in s \supset l \corners{= L}) \\
& & \wedge \forall u \forall z ((\S u, z) \in s \supset \exists v ((u,v) \in s \wedge z \corners{= M})))
\addtocounter{equation}{1} \end{eqnarray*}
\paragraph{Sets}
\begin{eqnarray*}
t \corners{\in X_i} & \equiv & t \in X_i \\
t \corners{\in \{ x : \N \mid P \}} & \equiv & [t/x] \corners{P}
\addtocounter{equation}{1} \end{eqnarray*}
\paragraph{Small Propositions}
\begin{eqnarray*}
\corners{M \hat{\eqN} N} & \equiv & \exists x (x \corners{= M} \wedge x \corners{= N}) \\
\corners{\hat{\bot}} & \equiv & \bot \\
\corners{P \hat{\supset} Q} & \equiv & \corners{P} \supset \corners{Q} \\
\corners{\hat{\forall} x : \N. P} & \equiv & \forall x \corners{P} \\
\corners{M \hat{\in}_\N N} & \equiv & \exists x (x \corners{= M} \wedge x \corners{\in N})
\addtocounter{equation}{1} \end{eqnarray*}
\paragraph{Propositions}
\begin{eqnarray*}
\corners{M \eqN N} & \equiv & \exists x (x \corners{= M} \wedge x \corners{= N}) \\
\corners{\bot} & \equiv & \bot \\
\corners{\phi \supset \psi} & \equiv & \corners{\phi} \supset \corners{\psi} \\
\corners{\forall x : \N. \phi} & \equiv & \forall x \corners{\phi} \\
\corners{\forall X : \SetN. \phi} & \equiv & \forall X \corners{\phi} \\
\corners{V(P)} & \equiv & \corners{P}
\addtocounter{equation}{1} \end{eqnarray*}
\caption{Interpretation of $\Tt$ in $\ACAO$}
\label{fig:TtACAO}
\end{figure}

\paragraph{Remark}
To interpret a term of the form $\R(L,[u,v]M,N)$, we make use of a standard technique for defining functions by recursion in $\ACAO$.  We are assuming we have defined in $\ACAO$ a pairing function $(m,n)$ on the natural numbers, and a coding of finite sequences of numbers as numbers, with $\Seq$ the set of all codes of sequences, and the formula $n \in s$ expressing that $n$ is a member of the sequence coded by $s$.  (For more details, see \cite[II.3]{simpson:sosoa}.)

Speaking informally, the formula $t \corners{= R(L,[u,v]M,N)}$ expresses that $(N,t)$ is a member of a sequence $s$, and that the members of this sequence $s$ must be
\[ (0, \R(L, [u,v]M, 0)), \quad (1, \R(L, [u,v]M, 1)), \quad \cdots, \quad (k, R(L,[u,v]M,k)) \]
up to some $k$, in some order.  It follows that $t = \R(L,[u,v]M,N)$.

The following theorem shows that the translation in Figure \ref{fig:TtACAO} is sound.
\begin{thm}[Soundness]
\label{thm:soundphi}
 \begin{enumerate}
  \item If $\Gamma \vdash M : \N$ then $\ACAO \vdash \exists ! x. x \corners{= M}$.
\item If $\Gamma \vdash M = M' : \N$ then $\ACAO \vdash \exists x (x \corners{= M} \wedge x \corners{= M'})$.
\item If $\Gamma \vdash M : \SetN$ then $\ACAO \vdash \exists ! X \forall x (x \in X \leftrightarrow x \corners{\in M})$.
\item If $\Gamma \vdash M = N : \SetN$ then $\ACAO \vdash \forall x (x \corners{\in M} \leftrightarrow x \corners{\in N})$.
\item If $\Gamma \vdash P = Q$ then $\ACAO \vdash \corners{P} \leftrightarrow \corners{Q}$.
\item If $\Gamma \vdash \phi = \psi$ then $\ACAO \vdash \corners{\phi} \leftrightarrow \corners{\psi}$.
\item If $\Gamma \vdash \phi_1, \ldots, \phi_n \Rightarrow \psi$ then $\ACAO \vdash \corners{\phi_1} \supset \cdots \supset \corners{\phi_n} \supset
\corners{\psi}$.
 \end{enumerate}
\end{thm}

\begin{pf}
We need the following two results first.
\begin{enumerate}
\item
For any term $M$ such that $x,y \notin \FV{M}$,
\[ \ACAO \vdash x \corners{= M} \supset y \corners{= M} \supset x = y \enspace . \]
This is proven by induction on $M$.
\item
Given a term $N$ such that $x \notin \FV{N}$, the following are all theorems of $\ACAO$:
\begin{gather}
 \label{eq:Phisub}
x \corners{= N} \supset (y \corners{=[N/x]M} \leftrightarrow y \corners{=M}) \\
x \corners{=N} \supset (y \corners{\in [N/x]M} \leftrightarrow y \corners{\in M}) \label{eq:Phisub2} \\
\forall x (x \in X \leftrightarrow x \corners{\in N}) \supset (y \corners{\in [N/X]M} \leftrightarrow y \corners{\in M}) \\
x \corners{= N} \supset (\corners{[N/x]P} \leftrightarrow \corners{P}) \\
\forall x (x \in X \leftrightarrow x \corners{\in N}) \supset (\corners{[N/X]P} \leftrightarrow \corners{P}) \label{eq:Phisub3} \\
x \corners{= N} \supset (\corners{[N/x]\phi} \leftrightarrow \corners{\phi}) \\
\forall x (x \in X \leftrightarrow x \corners{\in N}) \supset (\corners{[N/x]\phi} \leftrightarrow \corners{\phi})
\end{gather}
These are proven by induction on $M$, $P$ or $\phi$.  Formulas (\ref{eq:Phisub2})--(\ref{eq:Phisub3}) must be proven simultaneously.
\end{enumerate}
The seven parts of the theorem are now proven simultaneously by induction on derivations.  We deal with one case here: the rule
\[ \begin{prooftree}
    \begin{array}{c}
     \Gamma \vdash L : \N
\quad
\Gamma, u : \N, v : \N \vdash M : \N \\
\Gamma \vdash N : \N
    \end{array}
\justifies
\Gamma \vdash \R(L, [u,v]M, \s N) = [N/u, \R(L, [u,v]M, N)/v] M : \N
   \end{prooftree} \]
We reason in $\ACAO$.  By the induction hypothesis, there exist $l$ and $n$ such that $l \corners{= L}$, $n \corners{= N}$.  Further,
\[ \forall u \forall v \exists m . m \corners{= M} \enspace . \]
The following formula can be proven by induction on $z$:
\begin{alignat*}{1}
 \forall z \exists w \exists s \in \Seq & ((z,w) \in s \\
& \wedge \forall l( (0,l) \in s \supset l \corners{= L}) \\
& \wedge \forall u \forall z ((\S u, z) \in s \supset \exists v ((u,v) \in s \wedge z \corners{= M})))
\end{alignat*}
Now, let $n$ be the unique number such that $n \corners{= N}$.
There exist $m$, $p$ such that $(n,m)$ and $(\S n, p)$ are members of such a sequence $s$.  It follows that
\[ p \corners{= \R(L, [u,v]M, \s n)}, \; m \corners{= \R(L, [u,v]M, n)}, \; [n/u,m/v](p \corners{= M}) \enspace . \]
Hence, by (\ref{eq:Phisub}), we have
\[ p \corners{= \R(L, [u,v]M, \s N)}, \qquad p \corners{ = [N/u, \R(L, [u,v]M, N)/v]M} \]
as required.
\end{pf}

Conservativity shall follow from the following theorem, which states that the mapping $\corners{\,}$ is a left-inverse to the mapping $\angles{\,}$ from $\ACAO$
to $\Tt$, up to logical equivalence.

\begin{thm}$ $
\label{thm:linverse}
\begin{enumerate}
 \item For every term $t$ of $\ACAO$, we have $\ACAO \vdash t \corners{= \angles{t}}$.
\item For every arithmetic proposition $\phi$ of $\ACAO$, we have $\ACAO \vdash \phi \leftrightarrow \corners{\sangles{\phi}}$.
\item For every proposition $\phi$ of $\ACAO$, we have $\ACAO \vdash \phi \leftrightarrow \corners{\angles{\phi}}$.
\end{enumerate}
\end{thm}

\begin{pf}
 The proof of each of these statements is a straightforward induction.  We deal with one case here: the case $t \equiv t_1 + t_2$.  We reason in $\ACAO$.  The
induction hypothesis gives
\[ t_1 \corners{= \angles{t_1}}, \qquad t_2 \corners{= \angles{t_2}} \]
and we must show $t_1 + t_2 \corners{= \angles{t_1} \plus \angles{t_2}}$, i.e.
\begin{alignat*}{1}
 \exists n. \exists r \in \Seq & ( n \corners{= \angles{t_2}} \wedge (n, t_1 + t_2) \in r \\
& \wedge \forall l ((0,l) \in r \supset l \corners{= \angles{t_1}}) \\
& \wedge \forall x \forall z ((\S x,z) \in r \supset \exists y ((x,y) \in r \wedge z = \S y)))
\end{alignat*}
We prove the following by induction on $b$:
\begin{alignat*}{1}
\forall a,b .\exists r \in \Seq & ((b, a+b) \in r \\
& \wedge \forall l ((0,l) \in r \supset l = a) \\
& \wedge \forall x, z ((\S x,z) \in r \supset \exists y ((x,y) \in r \wedge z = \S y)))
\end{alignat*}
The desired proposition follows by instantiating $a$ with $t_1$ and $b$ with $t_2$.
\end{pf}

\begin{cor}[Conservativity of $\Tt$ over $\ACAO$]
\label{cor:TtACAO}
 For any formula $\phi$ of $\ACAO$, if $\Gamma \vdash \Rightarrow \angles{\phi}$ in $\Tt$, then $\ACAO \vdash \phi$.
\end{cor}

\begin{pf}
 By the Soundness Theorem, we have that $\ACAO \vdash \corners{\angles{\phi}}$.  By Theorem \ref{thm:linverse}, we have $\ACAO \vdash \phi \leftrightarrow
\corners{\angles{\phi}}$.  Therefore, $\ACAO \vdash \phi$.
\end{pf}

\section{Conservativity of $\LTTO$ over $\ACAO$}
\label{section:conservativity}

In this section, we shall prove that $\LTTO$ is conservative over $\Tt$.  This shall complete the proof that $\LTTO$ is conservative over $\ACAO$.

We shall do this by defining a number of subsystems of $\LTTO$ as shown in the diagram:
\[ \Tt \hookrightarrow \Tw \hookrightarrow \TwU \hookrightarrow \LTTO \enspace . \]
For each of these inclusions $A \hookrightarrow B$, we
shall prove that $A$ is a conservative subsystem of $B$; that is, for every judgement $\mathcal{J}$ in the language of $A$, if $\mathcal{J}$ is derivable in $B$
then $\mathcal{J}$ is derivable in $A$.  This shall sometimes involve constructing yet more subsystems in between $A$ and $B$, and proving that all these
inclusions are conservative.

Intuitively, each subsystem deals with a subset of the types of $\LTTO$.
\begin{itemize}
 \item $\Tt$ has only two types, $\N$ and $\Set{\N}$.
\item The types of $\Tw$ are all the types that can be built up from $\N$ using $\times$, $\rightarrow$ and $\Set{}$.
\item The types of $\TwU$ are the types of $\Tw$, together with the universe $U$.  (The constructors $\times$, $\rightarrow$ and $\Set{}$ may \emph{not} be
applied to $U$ in $\TwU$.)
\end{itemize}
The formal definitions of these systems shall be given in the sections to come.

\subsection{Digression --- Informal Explanation of Proof Technique}

Before proceeding with the technical details of the proof, we shall explain the informal ideas behind the technique we use to prove $\LTTO$ conservative over
$\Tt$.
The system $\LTTO$ is formed from $\Tt$ by adding products, function types, types of sets, and the universe $U$.  Intuitively, none of these should increase the
power
of the system.

We can see this most clearly in the case of products.  Speaking generally, let $S$ be any type system, and let $T$ be formed by adding product types to $S$. 
Then $T$ should have no more expressive power than $S$, because we can envisage a translation from $T$ to $S$:
\begin{itemize}
 \item wherever a variable $z : A \times B$ occurs, replace it with two variables \\ $x : A, y : B$;
\item wherever a term of type $A \times B$ occurs, replace it with two terms, one of type $A$ and one of type $B$.
\end{itemize}
As long as the only way of introducing terms of type $A \times B$ is the constructor $( \ , \ )$, we should always be able to find the two $S$-terms of types
$A$ and $B$ that correspond to
any $T$-term of type $A \times B$.
(This would however not be possible if (say) we could eliminate $\N$ over $A \times B$ in $T$.)

In brief:
\begin{itemize}
 \item the terms of type $A \times B$ can be interpreted as pairs $\langle M,N \rangle$ where $M : A$ and $N : B$.
\end{itemize}
Similarly,
\begin{itemize}
 \item the terms of type $A \rightarrow B$ can be interpreted as pairs $\langle x,M \rangle$ \\ where $x : A \vdash M : B$;
\item the terms of type $\Set{A}$ can be interpreted as pairs $\langle x, P \rangle$ \\ where $x : A \vdash P \p$.
\end{itemize}
Our proof relies on making these intuitive ideas formal.

These ideas show us how we might be able to remove types $A \rightarrow B$ that involve only one use of the arrow, but they do not show us how to handle types
of the form $(A \rightarrow B) \rightarrow C$.  Let us take another example: let $S$ be a typing system without function types, and let $T$ be formed from $S$
by adding function types.  Let us define the \emph{depth} of a type $A$, $d(A)$ by:
\begin{itemize}
 \item the depth of each type in $S$ is 0;
\item $d(A \rightarrow B) = \max(d(A),d(B)) + 1$.
\end{itemize}
Then we have seen how to interpret types of depth 1 in terms of types of depth 0.  More generally, we can interpret types of depth $n+1$ in terms of types of
depth $n$.

This shows us how to complete the proof. We introduce an infinite sequence of subsystems of $T$:
\[ S = \mathcal{A}_0 \hookrightarrow \mathcal{A}_1 \hookrightarrow \mathcal{A}_2 \hookrightarrow \cdots T \]
where, in $\mathcal{A}_n$, only types of depth $\leq n$ may occur.  We build an interpretation of $\mathcal{A}_{n+1}$ out of the terms of $\mathcal{A}_n$: every
type of $\mathcal{A}_n$ is interpreted as itself; the types $A \rightarrow B$ of depth $n+1$ are interpreted as the set of pairs $\langle x, M \rangle$ where $x
: A \vdash M : B$ in $\mathcal{A}_n$.

Using these interpretations, we can prove each $\mathcal{A}_{n+1}$ conservative over $\mathcal{A}_n$, and hence $T$ conservative over $S$.  With these intuitive
ideas to guide us, we return to the proof development.

\subsection{$\Tw$ is Conservative over $\Tt$}
\label{section:TwiTti}

We shall now define the system $\Tw$ to be $\Tt$ extended with pairs, functions and sets over all types, and prove that $\Tw$ is conservative over $\Tt$.

\begin{df}[$\Tw$]
 The LTT $\Tw$ is defined as follows.

The grammar of $\Tw$ is the grammar of $\Tt$ extended with
\[ \begin{array}{lrcl}
    \mbox{Type} & A & ::= & \cdots \mid A \times A \mid A \rightarrow A \mid \Set{A} \\
\mbox{Term} & M & ::= & \cdots \mid (M,M)_{A \times A} \mid \pi_1^{A \times A}(M) \mid \pi_2^{A \times A}(M) \mid \\
& & & \lambda x:A.M:A \mid M(M)_{A \rightarrow A} \mid \{ x:A \mid P \} \\
\mbox{Small Proposition} & P & ::= & \cdots \mid M \hat{\in}_A M
   \end{array} \]
The rules of deduction of $\Tw$ are the rules of deduction of $\Tt$, together with the rules for pairs (Appendix \ref{section:pairs}), function types (Appendix
\ref{section:functions}) and typed sets (Appendix \ref{section:sets}).
\end{df}

Note that the type-theory component $\Tw$ is non-dependent: a term can never occur in a type.  As a consequence, we have
\begin{lm}
 If $\Gamma \vdash A = B$ in $\Tw$ then $A \equiv B$.
\end{lm}

\begin{pf}
Induction on derivations.
\end{pf}

To prove that $\Tw$ is conservative over $\Tt$, we shall define an infinite sequence of subsystems of $\Tw$, and prove that each is conservative over the
previous subsystem, and that the smallest is conservative over $\Tt$.
\[ \Tt \hookrightarrow \mathcal{A}_1 \hookrightarrow \mathcal{A}_2 \hookrightarrow \cdots \Tw \]

We define the \emph{depth} of a type of $\Tw$ as follows.
\begin{df}
 Define the \emph{depth} $d(A) < \omega$ of a type $A$ of $\Tw$ by
\begin{eqnarray*}
 d(\N) & = & 0 \\
d(A \times B) & = & \max(d(A), d(B)) + 1 \\
d(A \rightarrow B) & = & \max(d(A), d(B)) + 1 \\
d(\Set{\N}) & = & 0 \\
d(\Set{A}) & = & d(A) + 1 & ($A \not\equiv \N$) \\
\addtocounter{equation}{1} \end{eqnarray*}
\end{df}

Note that the types of $\Tt$ are exactly the types of depth 0.

For $n \geq 1$, we shall define $\mathcal{A}_n$ to be the fragment of $\Tw$ that deals only with types of depth $\leq n$.
\begin{df}[$\mathcal{A}_n$]
 Let $n \geq 0$.  By a \emph{type} (term, small proposition, proposition, context, judgement) \emph{of $\mathcal{A}_n$}, we mean a type (term, small
proposition, proposition, context, judgement) of $\Tw$ that does not contain, as a subexpression, any type of depth $> n$.

We say a judgement $\mathcal{J}$ of $\mathcal{A}_n$ is \emph{derivable} in $\mathcal{A}_n$ iff there exists a derivation of $\mathcal{J}$ in $\Tw$ consisting
solely of judgements of $\mathcal{A}_n$; that is, a derivation of $\mathcal{J}$ in which no type of depth $> n$ occurs.  We write $\Gamma \vdash_n \mathcal{J}$
iff the judgement $\Gamma \vdash \mathcal{J}$ is derivable in $\mathcal{A}_n$.
\end{df}
Note that the types of $\mathcal{A}_n$ are exactly the types of depth $\leq n$.  Note also that $\mathcal{A}_0$ is just the system $\Tt$.

We shall prove that $\mathcal{A}_{n+1}$ is conservative over $\mathcal{A}_n$.  The proof shall involve defining an \emph{interpretation} of $\mathcal{A}_{n+1}$
in terms of the expressions of $\mathcal{A}_n$.  For the rest of this section, fix $n \geq 0$, and fix a context $\Delta$ of $\mathcal{A}_n$ such that $\Delta
\vdash_n \vald$.

\begin{df}[Interpretation of Types]
For the purposes of this definition, an `object' is either a term of $\mathcal{A}_n$, or a pair of terms of $\mathcal{A}_n$.

For every type $A$ of $\mathcal{A}_{n+1}$, we define the set of objects $\brackets{A}_\Delta$, and an equivalence relation $\sim_\Delta^A$ on this set, as
follows.

If $d(A) \leq n$, then
\begin{eqnarray*}
\brackets{A}_\Delta & = & \{  M \mid \Delta \vdash_n M : A \} \\
M \sim_\Delta^A N & \Leftrightarrow & \Delta \vdash_n M = N : A
\addtocounter{equation}{1} \end{eqnarray*}
Otherwise,
\begin{eqnarray*}
 \brackets{A \times B}_\Delta & = & \{ \langle M,N \rangle \mid \Delta \vdash_n M : A, \Delta \vdash_n N : B \} \\
\langle M,N \rangle \sim_\Delta^{A \times B} \langle M', N' \rangle & \Leftrightarrow & \Delta \vdash_n M = M' : A \wedge \Delta \vdash_n N = N' : B \\
& & \\
\brackets{A \rightarrow B}_\Delta & = & \{ \langle x,M \rangle \mid \Delta, x : A \vdash_n M : B \} \\
\langle x,M \rangle \sim_\Delta^{A \rightarrow B} \langle x, M' \rangle & \Leftrightarrow & \Delta, x : A \vdash_n M = M' : B \\
& & \\
\brackets{\Set{A}}_\Delta & = & \{ \langle x,P \rangle \mid \Delta, x : A \vdash_n P \p \} \\
\langle x,P \rangle \sim_\Delta^{\Set{A}} \langle x, P' \rangle & \Leftrightarrow & \Delta, x : A \vdash_n P = P'
\addtocounter{equation}{1} \end{eqnarray*}
We identify the elements of $\brackets{A \rightarrow B}_\Delta$ and $\brackets{\Set{A}}_\Delta$ up to $\alpha$-conversion; that is, we identify $\langle x,M
\rangle$ with $\langle y, [y/x]M \rangle$ if $y$ is not free in $M$.

We define the operations $\Pi_1$, $\Pi_2$ and $@$ on these objects as follows.
\begin{eqnarray*}
 \Pi_1( \langle M,N \rangle ) & \equiv & M \\
\Pi_2( \langle M,N \rangle ) & \equiv & N \\
\langle x,M \rangle @ N & \equiv & [N/x]M \\
\langle x,P \rangle @ N & \equiv & [N/x]P
\addtocounter{equation}{1} \end{eqnarray*}
$\Pi_1(X)$ and $\Pi_2(X)$ are undefined if $X$ is not a pair.  $X @ Y$ is undefined if $X$ does not have the form $\langle x,Z \rangle$, or if $Y$ is not a
term.
\end{df}
The intention is that we will interpret the terms of type $A$ as members of the set $\brackets{A}_\Delta$, with equal terms being interpreted as
$\sim_\Delta^A$-equivalent members

\begin{df}[Valuation]
Let $\Gamma \equiv x_1 : A_1, \ldots, x_n : A_n$ be a context of $\mathcal{A}_{n+1}$.  A \emph{$\Delta$-valuation} of $\Gamma$ is a function $v$ on $\{ x_1,
\ldots, x_n \}$ such that
\[ v(x_i) \in \brackets{A_i}_\Delta \qquad (i=1, \ldots, n) \enspace . \]
\end{df}

\begin{df}[Interpretation of Terms]
\label{df:interpA}
Given a term $M$ of $\mathcal{A}_{n+1}$ and a function $v$ whose domain includes $\FV{M}$, we define the object $\round{M}^v_A$ as follows.
\begin{eqnarray*}
 \round{x}^v & = & v(x) \\
\round{0}^v & \equiv & 0 \\
\round{\s M}^v & \simeq & \s \round{M}^v \\
\round{R(L, [x,y]M, N)}^v & \simeq & R(\round{L}^v, [x,y]\round{M}^{v[x:=x,y:=y]}, \round{N}^v) \\
\round{(M,N)_{A \times B}}^v & \simeq & \begin{cases}
	(\round{M}^v, \round{N}^v)_{A \times B} & \mbox{if } d(A \times B) \leq n \\
	\langle \round{M}^v, \round{N}^v \rangle & \mbox{if } d(A \times B) = n + 1
	\end{cases} \\
\round{(\pi_1^{A \times B}(M))}^v & \simeq & \begin{cases}
	\pi_1^{A \times B}(\round{M}^v) & \mbox{if } d(A \times B) \leq n \\
	\Pi_1(\round{M}^v) & \mbox{if } d(A \times B) = n + 1
                                \end{cases}
\addtocounter{equation}{1} \end{eqnarray*}
\begin{eqnarray*}
\round{(\pi_2^{A \times B}(M))}^v & \simeq & \begin{cases}
                                 \pi_2^{A \times B}(\round{M}^v) & \mbox{if } d(A \times B) \leq n\\
\Pi_2(\round{M}^v) & \mbox{if } d(A \times B) = n + 1
                                \end{cases} \\
\round{\lambda x:A.M:B}^v & \simeq & \begin{cases}
                                    \lambda x:A.\round{M}^{v[x:=x]}:B & \mbox{if } d(A \rightarrow B) \leq n \\
\langle x, \round{M}^{v[x:=x]} \rangle & \mbox{if } d(A \rightarrow B) = n + 1
                                   \end{cases} \\
\round{M(N)_{A \rightarrow B}}^v & \simeq & \begin{cases}
                           \round{M}^v(\round{N}^v)_{A \rightarrow B} & \mbox{if } d(A \rightarrow B) \leq n \\
\round{M}^v @ \round{N}^v & \mbox{if } d(A \rightarrow B) = n + 1
                          \end{cases} \\
\round{\{x : A \mid P \}}^v & \simeq & \begin{cases}
\{ x : A \mid \round{P}^{v[x:=x]} \} & \mbox{if } d(\Set{A}) \leq n \\
\langle x, \round{P}^{v[x:=x]} \rangle & \mbox{if } d(\Set{A}) = n + 1
\end{cases}
\addtocounter{equation}{1} \end{eqnarray*}
\end{df}
Note that this is a \emph{partial} definition; $\round{M}^v_A$ will sometimes be undefined.

\begin{df}[Interpretation of Small Propositions]
If $P$ is a small \\ $\mathcal{A}_{n+1}$-proposition, we define the small proposition $\round{P}^v$ of $\mathcal{A}_n$.
\begin{eqnarray*}
\round{M \hat{\eqN} N}^v & \simeq & \round{M}^v \hat{\eqN} \round{N}^v \\
\round{\hat{\bot}}^v & \equiv & \hat{\bot} \\
\round{P \hat{\supset} Q}^v & \simeq & \round{P}^v \hat{\supset} \round{Q}^v \\
\round{\hat{\forall} x : \N. P}^v & \simeq & \hat{\forall} x:\N.\round{P}^{v[x:=x]} \\
\round{M \hat{\in}_A N}^v & \simeq & \begin{cases}
                              \round{M}^v \hat{\in}_A \round{N}^v & \mbox{if } d(A) \leq n \\
\round{N}^v @ \round{M}^v & \mbox{if } d(A) = n + 1
                             \end{cases}
\addtocounter{equation}{1} \end{eqnarray*}
\end{df}

\begin{df}[Depth of a Proposition]
We define the \emph{depth} of a proposition $\phi$, $d(\phi)$, to be
\[
d(\phi) = \begin{cases}
0 & \mbox{if } \phi \mbox{ is quantifier-free} \\
\max\{d(A) \mid \phi \mbox{ contains a quantifier } \forall x:A \} & \mbox{otherwise}
\end{cases} \]
\end{df}

\begin{df}[Interpretation of Propositions]
If $\phi$ is \\ an $\mathcal{A}_{n+1}$-proposition of depth $\leq n$, we define the $\mathcal{A}_n$-proposition $\round{\phi}^v$ as follows
\begin{eqnarray*}
\round{M \eqN N}^v & \simeq & \round{M}^v \eqN \round{N}^v \\
\round{\bot}^v & \equiv & \bot \\
\round{\phi \supset \psi}^v & \simeq & \round{\phi}^v \supset \round{\psi}^v \\
\round{\forall x:A.\phi}^v & \simeq & \forall x:A.\round{\phi}^{v[x:=x]} \\
\round{V(P)}^v & \simeq & V(\round{P}^v)
\addtocounter{equation}{1} \end{eqnarray*}
\end{df}

We have defined a sound interpretation of all the judgement forms of $\mathcal{A}_{n+1}$ except one: the judgement form $\Gamma \vdash \Phi \Rightarrow \phi$. 
To interpret these judgements, we shall define a notion of \emph{satisfaction}.  Intuitively, we define what it is for a proposition $\phi$ of
$\mathcal{A}_{n+1}$ to be `true' under a context $\Delta$, valuation $v$ and sequence of propositions $\Phi$ of $\mathcal{A}_n$.

\begin{df}[Satisfaction]
\label{df:satis}
Let $\Phi \equiv \phi_1, \ldots, \phi_m$ be a sequence of propositions of $\mathcal{A}_n$
such that $\Delta \vdash_n \phi_1 \prop$, \ldots, $\Delta \vdash_n \phi_m \prop$.  Let $v$ be a $\Delta$-valuation of $\Gamma$.  Suppose $\Gamma \vdash \phi
\prop$.  We define what it means for $(\Delta, \Phi, v)$ to \emph{satisfy} $\phi$, $(\Delta, \Phi, v) \models \phi$, as follows.

If $d(\phi) \leq n$, then
$((\Delta, \Phi, v) \models \phi) \Leftrightarrow (\Delta \vdash_n \Phi \Rightarrow \round{\phi}^v)$.

Otherwise,
\begin{itemize}
\item
$(\Delta, \Phi, v) \models \phi \supset \psi$ iff, for all $\Delta' \supseteq \Delta$ and $\Phi' \supseteq \Phi$, if $(\Delta', \Phi', v) \models \phi$ then
$(\Delta', \Phi', v) \models \psi$.
\item
$(\Delta, \Phi, v) \models \forall x:A.\phi$ iff, for all $\Delta' \supseteq \Delta$ and $a \in \brackets{A}_{\Delta'}$, we have \\ $(\Delta', \Phi, v[x:=a])
\models \phi$.
\end{itemize}
\end{df}

\begin{df}[Satisfaction and Truth]
Let $\Gamma \vdash \mathcal{J}$ be a judgement of $\mathcal{A}_{n+1}$, and let $v$ be a $\Delta$-valuation of $\Gamma$.  We define what it means for $\Delta$
and $v$ to \emph{satisfy} $\mathcal{J}$, written $(\Delta,v) \models \mathcal{J}$, as follows:
\begin{itemize}
\item
$(\Delta, v) \models M : A$ iff $\round{M}^v \in \brackets{A}_\Delta$.
\item
$(\Delta, v) \models M = N : A$ iff $\round{M}^v \sim_\Delta^A \round{N}^v$.
\item
$(\Delta, v) \models P \p$ iff $\Delta \vdash_n \round{P}^v \p$.
\item
$(\Delta, v) \models P = Q$ iff $\Delta \vdash_n \round{P}^v = \round{Q}^v$.
\item
If $d(\phi) \leq n$, then $(\Delta, v) \models \phi \prop$ iff $\Delta \vdash_n \round{\phi}^v \prop$.
\item
$(\Delta, v) \models \phi = \psi$ iff for all $\Phi$, $(\Delta, \Phi, v) \models \phi \Leftrightarrow (\Delta, \Phi, v) \models \psi$.
\item
$(\Delta, v) \models \psi_1, \ldots, \psi_n \Rightarrow \chi$ iff, for all $\Phi$, if $(\Delta, \Phi, v) \models \psi_i$ for $1 \leq i \leq n$ then $(\Delta,
\Phi, v) \models \chi$.
\item
For all other judgement bodies $\mathcal{J}$, we have $(\Delta, v) \models \mathcal{J}$ for all $\Delta$, $v$.
\end{itemize}
We say a judgement $\Gamma \vdash \mathcal{J}$ of $\mathcal{A}_{n+1}$ is \emph{true} iff, for every context $\Delta$ of $\mathcal{A}_n$ such that $\Delta
\vdash_n \vald$ and every $\Delta$-valuation $v$ of $\Gamma$, $(\Delta, v) \models \mathcal{J}$.
\end{df}

The following theorem shows that this interpretation is sound.
\begin{thm}[Soundness]
\label{thm:soundness}
Every derivable judgement of $\mathcal{A}_{n+1}$ is true.
\end{thm}

The proof is given in Appendix \ref{app:pfsound}.

\begin{thm}[Completeness]$ $
\label{thm:completeness}
\begin{enumerate}
\item
Let $\Gamma \vdash \mathcal{J}$ be a judgement of $\mathcal{A}_n$, and suppose $\mathcal{J}$ does not have the form $\Phi \Rightarrow \psi$.
If the judgement is true, and $\Gamma \vdash_n \vald$, then the judgement is derivable in $\mathcal{A}_n$.
\item
Let $\Gamma \vdash \phi_1, \ldots, \phi_m \Rightarrow \psi$ be a judgement of $\mathcal{A}_n$.  If the judgement is true, and we have $\Gamma \vdash_n \vald$
and $\Gamma \vdash_n \phi_i \prop$ for $i=1, \ldots, m$, then the judgement is derivable in $\mathcal{A}_n$.
\end{enumerate}
\end{thm}

\begin{pf}$ $
\begin{enumerate}
\item
Let $1_\Gamma$ be the identity function on $\dom \Gamma$.
Then $1_\Gamma$ is a $\Gamma$-valuation of $\Gamma$ and, for every expression $X$ of $\mathcal{A}_n$ such that $\FV{X} \subseteq \dom \Gamma$,
\[ \round{X}^{1_\Gamma} \equiv X \enspace . \]

So, suppose $\Gamma \vdash M : A$ is a judgement of $\mathcal{A}_n$, and is true.  Then
\[ (\Gamma, 1_\Gamma) \models M : A \]
and so $\Gamma \vdash_n \round{M}^{1_\Gamma} : A$.  But $\round{M}^{1_\Gamma} \equiv M$, and so $\Gamma \vdash_n M : A$ as required.

The proof for the other judgement forms is similar.
\item
Suppose $\Gamma \vdash \Phi \Rightarrow \psi$ is true, where $\Phi \equiv \phi_1, \ldots, \phi_m$.  We have that
\[ \Gamma \vdash \Phi \Rightarrow \phi_i \qquad (i=1, \ldots, m) \]
and so $(\Gamma, \Phi, 1_\Gamma)$ satisfies each $\phi_i$.  Therefore, $(\Gamma, \Phi, 1_\Gamma)$ satisfies $\psi$, that is
\[ \Gamma \vdash \Phi \Rightarrow \psi \]
as required.
\end{enumerate}
\end{pf}

\begin{cor}
\label{cor:cons}
 If $\mathcal{J}$ is a judgement of $\mathcal{A}_n$ derivable in $\mathcal{A}_{n+1}$, then $\mathcal{J}$ is derivable in $\mathcal{A}_n$.
\end{cor}

\begin{pf}
This follows almost immediately from the Soundness Theorem and the Completeness Theorem.  There are just two facts that need to be verified:
\begin{enumerate}
\item
If $\Gamma$ is a context of $\mathcal{A}_n$, and $\Gamma \vdash_{n+1} \mathcal{J}$, then $\Gamma \vdash_n \vald$.
\item
If $\Gamma$ is a context of $\mathcal{A}_n$; $\phi_1$, \ldots, $\phi_m$ are propositions of $\mathcal{A}_n$; and $\Gamma \vdash_{n+1} \phi_1, \ldots, \phi_m
\Rightarrow \psi$; then $\Gamma \vdash_n \vald$ and $\Gamma \vdash_n \phi_i \prop$.
\end{enumerate}
These are proven fairly easily by induction on derivations, using the Soundness and Completeness Theorems.
\end{pf}

\begin{cor}[Conservativity of $\Tw$ over $\Tt$]
\label{cor:TwTt}
If $\mathcal{J}$ is a judgement of $\Tt$, and $\mathcal{J}$ is derivable in $\Tw$, then $\mathcal{J}$ is derivable in $\Tt$.
\end{cor}

\begin{pf}
 Suppose $\mathcal{J}$ is derivable in $\Tw$.  Let $n$ be the largest depth of type or proposition that occurs in the derivation.  Then $\mathcal{J}$ is
derivable in $\mathcal{A}_n$.  Applying Corollary \ref{cor:cons}, we have that $\mathcal{J}$ is derivable in $\mathcal{A}_{n-1}$, $\mathcal{A}_{n-2}$, \ldots,
$\mathcal{A}_0$.  But derivability in $\mathcal{A}_0$ is the same as derivability in $\Tt$.
\end{pf}

\subsection{$\TwU$ is Conservative over $\Tw$}
\label{section:TwUTw}
The system $\TwU$ is the fragment of $\LTTO$ that includes all the types of $\Tw$, and the universe $U$, but does not include types such as $U \times U$, $\N
\rightarrow U$, or $\Set{U}$.  It is defined in a similar manner to the systems $\mathcal{A}_n$ of the previous section, but using a new notion of depth.

\begin{df}[$\TwU$]
A type $A$ of $\LTTO$ is a \emph{type of $\TwU$}, iff either $A \equiv U$ or the symbol $U$ does not occur in $A$.

By a \emph{term}
(small proposition, proposition, context, judgement) of $\TwU$, we mean a term (small proposition, proposition, context, judgement) of $\LTTO$ in which every
type that occurs as a subexpression is a type of $\TwU$

We say a judgement $\mathcal{J}$ of $\TwU$ is \emph{derivable} in $\TwU$ iff there exists a derivation of $\mathcal{J}$ in $\LTTO$ consisting solely of
judgements of $\TwU$; that is, a derivation of $\mathcal{J}$ in which every type that occurs is a type of $\TwU$.

We write $\Gamma \vdashu \mathcal{J}$ iff the judgement $\Gamma \vdash \mathcal{J}$ is derivable in $\TwU$, and $\Gamma \vdashw \mathcal{J}$ iff the judgement
$\Gamma \vdash \mathcal{J}$ is derivable in $\Tw$.
\end{df}

\paragraph{Note}
The types of $\TwU$ are not closed under $\times$, $\rightarrow$ or $\Set{\,}$.  For example, the types $U \times U$ and $U \rightarrow U$ are not types of
$\TwU$. 

In order to prove $\TwU$ conservative over $\Tw$, we must find an interpretation of $U$ and of the types $T(M)$.  We do this by interpreting the objects of
$T(M)$ as \emph{binary trees} with leaves labelled by natural numbers.  For example, the object $((1,2),3)$ of type $T((\hat{\N} \hat{\times} \hat{\N})
\hat{\times} \hat{\N})$ will be interpreted as the binary tree
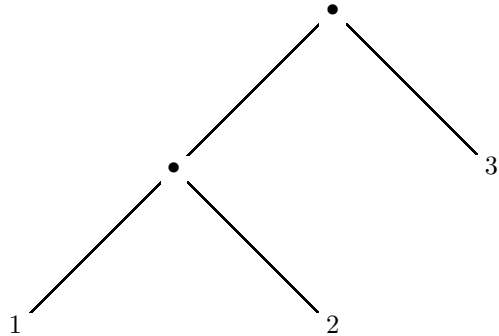
\[ \begin{diagram}
& & & & \bullet & & \\
& & & \ldLine & & \rdLine & \\
& & \bullet & & & & 3 \\
& \ldLine & & \rdLine & & & \\
1 & & & & 2 & &
\end{diagram} \]
We interpret $U$ as the set of all \emph{shapes} of binary tree.  We begin by inventing a syntax for the set of all shapes of binary trees:
\begin{df}[Shape]
 The set of \emph{shapes} is defined inductively by:
\begin{itemize}
 \item $\bullet$ is a shape.
\item If $S$ and $T$ are shapes, so is $S \wedge T$.
\end{itemize}
We write $\mathscr{S}$ for the set of all shapes.
\end{df}
The example tree above has shape $(\bullet \wedge \bullet) \wedge \bullet$.

We must thus associate each shape with a small type.  This association is done formally by the following function:
\begin{df}
 For every shape $S \in \mathscr{S}$, define the type $\TT(S)$ of $\Tw$ as follows:
\begin{eqnarray*}
 \TT(\bullet) & \equiv & \N \\
\TT(S \wedge T) & \equiv & \TT(S) \times \TT(T) \enspace .
\addtocounter{equation}{1} \end{eqnarray*}
\end{df}

There are two other gaps between $\TwU$ and $\Tw$ to be bridged.  In $\Tw$, we can only eliminate $\N$ over $\N$; in $\TwU$, we can eliminate over any small
type.  Likewise, in $\Tw$, a small proposition may only involve quantification over $\N$; in $\TwU$, a small proposition may involve quantification over any
small type.

We bridge these gaps by using the fact that every binary tree can be \emph{coded} as a natural number.  Given a bijection $P : \N^2 \rightarrow \N$, we can
assign a code number to every binary tree.  The binary tree above, for example, would be assigned the code number $P(P(1,2),3)$.  We shall define, for every
shape $S$, mutually inverse functions
\begin{eqnarray*}
 \code_S & : & \TT(S) \rightarrow \N \\
\decode_S & : & \N \rightarrow \TT(S)
\addtocounter{equation}{1} \end{eqnarray*}
Using these functions, we can interpret recursion over small types by recursion over $\N$, and quantification over small types by quantification over $\N$.

We turn now to the formal details.  The first step is to construct in $\Tw$ the bijection $P$ above, and the coding and decoding functions.
\begin{lm}[Pairing Function]
\label{lm:pairing}
 There exist $\Tw$-terms
\begin{eqnarray*}
 \PP & : & \mathbb{N} \times \mathbb{N} \rightarrow \mathbb{N} \\
\QQ & : & \mathbb{N} \rightarrow \mathbb{N} \\
\RR & : & \mathbb{N} \rightarrow \mathbb{N}
\addtocounter{equation}{1} \end{eqnarray*}
such that the following are theorems of $\Tw$:
\begin{equation}
\label{eq:PPQQRR}
\left. \begin{array}{c}
 \forall x : \mathbb{N}. \forall y : \mathbb{N}. \QQ(\PP(x,y)) =_\mathbb{N} x \\
\forall x : \mathbb{N}. \forall y : \mathbb{N}. \RR(\PP(x,y)) =_\mathbb{N} y \\
\forall x : \mathbb{N}. x =_\N \PP(\QQ(x), \RR(x))
\end{array} \right\}
\end{equation}
\end{lm}

\begin{pf}
Consider the three primitive recursive functions
\begin{eqnarray*}
p(m,n) & = & 2^m(2n+1) \\
q(n) & = & \mbox{the greatest } m \mbox{ such that } 2^m \mbox{ divides } n \\
r(n) & = & 1/2 ( n/2^{q(n)} - 1 )
\addtocounter{equation}{1} \end{eqnarray*}
It is straightforward to define terms $\PP$, $\QQ$ and $\RR$ in $\Tw$ that express $p$, $q$ and $r$ and prove the three formulas (\ref{eq:PPQQRR}).
\end{pf}

Fix three such terms $\PP$, $\QQ$ and $\RR$ for the sequel.

We shall also need a notion of equality on every small type in $\Tw$, not just $\N$.  This is defined as follows.
\begin{df}
\label{df:propeq}
Given $\Tw$-terms $M$ and $N$ and a $\Tw$-type $A$, define \\ the $\Tw$-proposition $M =_A N$ as follows.
\begin{eqnarray*}
 M =_{\N} N & \equiv & M \eqN N \\
M =_{A \times B} N & \equiv & \pi_1(M) =_A \pi_1(N) \wedge \pi_2(M) =_B \pi_2(N) \\
M =_{A \rightarrow B} N & \equiv & \forall x :A. M(x) =_B N(x) \\
M =_{\Set{A}} N & \equiv & \forall x : A.( x \in_A M \leftrightarrow x \in_A N)
\addtocounter{equation}{1} \end{eqnarray*}
\end{df}

\begin{df}[Coding Functions]
For each shape $S \in \mathscr{S}$, define \\ the $\Tw$-terms
\begin{eqnarray*}
 \code_S & : & \TT(S) \rightarrow \N \\
\decode_S & : & \N \rightarrow \TT(S)
\addtocounter{equation}{1} \end{eqnarray*}
as follows.
\begin{eqnarray*}
 \code_\bullet & \equiv & \lambda x:\N.x \\
\decode_\bullet & \equiv & \lambda x:\N.x \\
& & \\
\code_{S \wedge T} & \equiv & \lambda p:\TT(S) \times \TT(T). \PP(\code_S(\pi_1(p)), \code_T(\pi_2(p))) \\
\decode_{S \wedge T} & \equiv & \lambda n:\N. (\decode_S(\QQ(n)), \decode_T(\RR(n)))
\addtocounter{equation}{1} \end{eqnarray*}
\end{df}

\begin{lm}
\label{lm:codedecode}
 For every shape $S$, the following are theorems of $\Tw$:
\begin{gather*}
 \forall p : \TT(S). \decode_S(\code_S(p)) =_{\TT(S)} p \\
\forall n : \N. \code_S(\decode_S(n)) =_{\N} n
\end{gather*}
\end{lm}

\begin{pf}
 The proof is by induction on $S$, using the properties of $\PP$, $\QQ$ and $\RR$ from Lemma \ref{lm:pairing}.
\end{pf}

We can now proceed to define our interpretation of $\TwU$ in terms of $\Tw$.  The definition is more complex than the interpretation in the previous section,
because the type-theoretic component of $\TwU$ is dependent, so we must define our interpretations of terms and types simultaneously.
\begin{df}
\label{df:interpTwU}
Let $\Delta$ be a context of $\Tw$, and $v$ a function.  We define the following simultaneously.
\begin{itemize}
\item
Given a $\TwU$-term $M$ and a function $v$, define the object $\round{M}^v$ as follows.
\begin{eqnarray*}
 \round{x}^v & \simeq & v(x) \\
\round{0}^v & \equiv & 0 \\
\round{\s M}^v & \simeq & \s \round{M}^v \\
\round{(M,N)_{A \times B}}^v & \simeq & (\round{M}^v, \round{N}^v)_{\round{A}^v \times \round{B}^v} \\
\round{\pi_1^{A \times B}(M)}^v & \simeq & \pi_1^{\round{A}^v \times \round{B}^v}(\round{M}^v) \\
\round{\pi_2^{A \times B}(M)}^v & \simeq & \pi_2^{\round{A}^v \times \round{B}^v}(\round{M}^v) \\
\round{\lambda x:A.M:B}^v & \simeq & \lambda x:\round{A}^v.\round{M}^{v[x:=x]}:\round{B}^{v} \\
\round{M(N)_{A \rightarrow B}}^v & \simeq & \round{M}^v (\round{N}^v)_{\round{A}^v \rightarrow \round{B}^v} \\
\round{\hat{\N}}^v & = & \bullet \\
\round{M \hat{\times} N}^v & \simeq & \round{M}^v \wedge \round{N}^v \\
\round{\{ x : A \mid P \}}^v & \simeq & \{ x : \round{A}^v \mid \round{P}^{v[x:=x]} \} \\
\round{\EN([x]T(K),L,[x,y]M,N)}^v & \simeq & \decode_{S(\round{N}^v)} (\R (\code_{S(0)}(\round{L}^v), \\
& & \quad  
[x,y] \code_{S(\s x)}(\round{M}^{v'}), \round{N}^v))
\addtocounter{equation}{1} \end{eqnarray*}
where
$S(N) \equiv \round{K}^{v[x:=N]}$ and
$v' = v[x:=x,y:=\decode_{S(x)}(y)]$.

\item Given a type $A \not\equiv U$ of $\TwU$, define a type $\round{A}^v$ of $\Tw$.
\begin{eqnarray*}
 \round{\N}^v & \equiv & \N \\
\round{A \times B}^v & \simeq & \round{A}^v \times \round{B}^v \\
\round{A \rightarrow B}^v & \simeq & \round{A}^v \rightarrow \round{B}^v \\
\round{T(M)}^v & \simeq & T(\round{M}^v) \\
\round{\Set{A}}^v & \simeq & \Set{\round{A}^v}
\addtocounter{equation}{1} \end{eqnarray*}
\item Given a $\TwU$-type $A$, define a set $\brackets{A}_\Delta^v$ and an equivalence relation $\sim_{\Delta v}^A$ on $\brackets{A}_\Delta^v$ as follows.

If $A \not\equiv U$, then
\begin{eqnarray*}
 \brackets{A}^v & = & \{ M \mid \Delta \vdashu M : \round{A}^v \} \\
M \sim_{\Delta v}^A N & \Leftrightarrow & \Delta \vdashu \Rightarrow M =_{\round{A}^v} N
\addtocounter{equation}{1} \end{eqnarray*}
Otherwise,
\begin{eqnarray*}
\brackets{U}^v & = & \mathscr{S} \\
S \sim_{\Delta v}^U T & \Leftrightarrow & S = T
\addtocounter{equation}{1} \end{eqnarray*}
\item
Let $\Gamma \equiv x_1 : A_1, \ldots, x_m : A_m$ be a context of $\TwU$.
We say that $v$ is a \emph{$\Delta$-valuation} of $\Gamma$ iff $v(x_i) \in \brackets{A_i}_\Delta^v$ for $i=1, \ldots, n$.
\item
Given a small proposition $P$ of $\TwU$, define a small proposition $\round{P}^v$ of $\Tw$ as follows.
\begin{eqnarray*}
\round{M_1 \hat{=}_N M_2}^v & \simeq & \code_{\round{N}^v} (\round{M_1}^v) \hat{\eqN} \code_{\round{N}^v}(\round{M_2}^v) \\
\round{\hat{\bot}}^v & \equiv & \hat{\bot} \\
\round{P \hat{\supset} Q}^v & \equiv & \round{P}^v \hat{\supset} \round{Q}^v \\
\round{\hat{\forall} x : M.P}^v & \simeq & \hat{\forall} x : \N. \round{P}^{v[x:= \decode_{\round{M}^v}(x)]} \\
\round{M \hat{\in}_A N}^v & \simeq & \round{M}^v \hat{\in}_{\round{A}^v} \round{N}^v
\addtocounter{equation}{1} \end{eqnarray*}
\item
Given a proposition $\phi$ of $\TwU$ that does not include a quantifier over $U$, define a proposition $\round{\phi}^v$ of $\Tw$ as follows.
\begin{eqnarray*}
 \round{M_1 =_N M_2}^v & \simeq & \round{M_1}^v =_{\TT(\round{N}^v)} \round{M_2}^v \\
\round{\bot}^v & \equiv & \bot \\
\round{\phi \supset \psi}^v & \simeq & \round{\phi}^v \supset \round{\psi}^v \\
\round{\forall x:A.\phi}^v & \simeq & \forall x : \round{A}^v . \round{\phi}^{v[x:=x]} \\
\round{V(P)}^v & \simeq & V(\round{P}^v)
\addtocounter{equation}{1} \end{eqnarray*}
\end{itemize}
\end{df}

Recall that we write $\Delta \vdashw \mathcal{J}$ iff $\Delta \vdash \mathcal{J}$ is derivable in $\Tw$.

\begin{df}[Satisfaction]
Let $\Phi \equiv \phi_1, \ldots, \phi_m$ be a sequence of propositions of $\Tw$ such that $\Delta \vdashw \phi_i \prop$.  Let $v$ be a $\Delta$-valuation of
$\Gamma$.  Suppose $\Gamma \vdash \phi \prop$.  We define what it means for $(\Delta, \Phi, v)$ to \emph{satisfy} $\phi$, $(\Delta, \Phi, v) \models \phi$, as
follows.

If $\phi$ does not involve quantification over $U$, then
\[ ((\Delta, \Phi, v) \models \phi) \Leftrightarrow (\Delta \vdashw \Phi \Rightarrow \round{\phi}^v) \enspace . \]

Otherwise,
\begin{itemize}
\item $(\Delta, \Phi, v) \models \phi \supset \psi$ iff, for all $\Delta' \supseteq \Delta$ and $\Phi' \supseteq \Phi$, if $(\Delta', \Phi', v) \models \phi$
then $(\Delta', \Phi', v) \models \psi$.
\item $(\Delta, \Phi, v) \models \forall x : A. \phi$ iff, for all $\Delta' \supseteq \Delta$ and $a \in \brackets{A}_\Delta^v$, we have \\ $(\Delta', \Phi,
v[x:=a]) \models \phi$.
\end{itemize}
\end{df}

\begin{df}[Satisfaction and Truth]
Let $\Gamma \vdash \mathcal{J}$ be a judgement of $\TwU$.  Let $\Delta \vdashw \vald$, and let $v$ be a $\Delta$-valuation of $\Gamma$.  We define what it means
for $\Delta$ and $v$ to \emph{satisfy} $\mathcal{J}$, $(\Delta, v) \models \mathcal{J}$, as follows.
\begin{itemize}
\item
If $A \not\equiv U$, then $(\Delta, v) \models A \type$ iff $\round{A}^v$ is defined.
\item
If $A \not\equiv U \not\equiv B$, then $(\Delta, v) \models A = B$ iff $\round{A}^v \equiv \round{B}^v$.
\item 
$(\Delta, v) \models M : A$ iff $\round{M}^v \in \brackets{A}_\Delta^v$.
\item
$(\Delta, v) \models M = N : A$ iff $\round{M}^v \sim_{\Delta v}^A \round{N}^v$.
\item
$(\Delta ,v) \models P \p$ iff $\Delta \vdashw \round{P}^v \p$.
\item
$(\Delta, v) \models P = Q$ iff $\Delta \vdashw \Rightarrow V(\round{P}^v) \leftrightarrow V(\round{Q}^v)$.
\item
If $\phi$ does not include a quantifier over $U$, then $(\Delta, v) \models \phi \prop$ iff \\ $\Delta \vdashw \round{\phi} \prop$.
\item
$(\Delta, v) \models \phi = \psi$ iff, for all $\Phi$, we have $(\Delta, \Phi, v) \models \phi$ iff $(\Delta, \Phi, v) \models \psi$.
\item
$(\Delta, v) \models \phi_1, \ldots, \phi_m \Rightarrow \psi$ iff, for all $\Phi$, if $(\Delta, \Phi, v) \models \phi_i$ for $i=1, \ldots, m$ then $(\Delta,
\Phi, v) \models \psi$.
\item
For all other judgement forms, we have $(\Delta, v) \models \mathcal{J}$ for all $\Delta$, $v$.
\end{itemize}
We say a judgement $\Gamma \vdash \mathcal{J}$ of $\TwU$ is \emph{true} iff, for all $\Delta$ such that $\Delta \vdashw \vald$ and all $\Delta$-valuations $v$
of $\Gamma$, $(\Delta, v) \models \mathcal{J}$.
\end{df}
\nopagebreak
\paragraph{Remark}
This interpretation uses the propositional equality defined in Definition \ref{df:propeq}, whereas our interpretation in the previous section used judgemental
equality.
This is because the properties of our coding and decoding functions can be shown to hold up to propositional equality (as in Lemma \ref{lm:codedecode}), but not
up to judgemental equality.

We now prove that the interpretation is sound.
\begin{thm}[Soundness]
\label{thm:soundness'}
Every derivable judgement in $\TwU$ is true.
\end{thm}

The proof is given in Appendix \ref{app:pfsound'}.

\begin{thm}[Completeness]
\label{thm:completeness2}
If $\Gamma \vdash \mathcal{J}$ is a judgement of $\Tw$ that is true, and $\Gamma \vdashw \vald$, then $\Gamma \vdash \mathcal{J}$ is derivable in $\Tw$.
\end{thm}

\begin{pf}
Exactly as in Theorem \ref{thm:completeness}.
\end{pf}

\begin{cor}
\label{cor:TwUTw}
 If $\mathcal{J}$ is a judgement of $\Tw$ derivable in $\TwU$, then $\mathcal{J}$ is derivable in $\Tw$.
\end{cor}

\begin{pf}
Similar to Corollary \ref{cor:cons}.
\end{pf}

\subsection{$\LTTO$ is Conservative over $\TwU$}
\label{section:LTTOTwU}
The next step in our proof is to apply the same method to show that $\LTTO$ is conservative over $\TwU$.  The proof is very similar to Section
\ref{section:TwiTti}, but the details are more complicated, because we are now dealing with LTTs whose type theoretic components use dependent types.

Once again, we introduce an infinite sequence of subsystems between $\TwU$ and $\LTTO$:
\[ \TwU = \mathcal{B}_0 \hookrightarrow \mathcal{B}_1 \hookrightarrow \mathcal{B}_2 \hookrightarrow \cdots \LTTO \]

We do this using a new definition of the depth of a type:
\begin{df}[Depth]
Define the \emph{depth} $D(A)$ of a type $A$ of $\LTTO$ by
\begin{eqnarray*}
 D(\N) & = & 0 \\
D(A \times B) & = & \begin{cases}
                     0 & \mbox{if } D(A) = D(B) = 0 \\
\max(D(A), D(B)) + 1 & \mbox{otherwise}
                    \end{cases} \\
D(A \rightarrow B) & = & \begin{cases}
                          0 & \mbox{if } D(A) = D(B) = 0 \\
\max(D(A), D(B)) + 1 & \mbox{otherwise}
                         \end{cases} \\
D(\Set{A}) & = & \begin{cases}
                  0 & \mbox{if } D(A) = 0 \\
D(A) + 1 & \mbox{otherwise}
                 \end{cases} \\
D(U) & = & 1 \\
D(T(M)) & = & 0
\addtocounter{equation}{1} \end{eqnarray*}
We define the depth of a proposition $\phi$, $D(\phi)$, to be the largest depth of a type $A$ such that the quantifier $\forall x:A$ occurs in $\phi$, or
$D(\phi) = 0$ if $\phi$ is quantifier-free.
\end{df}
Note that the types of $\TwU$ are exactly the types $A$ such that $D(A) \leq 1$.

The subsystems $\mathcal{B}_n$ are defined as follows.
\begin{df}[$\mathcal{B}_n$]
 Let $n \geq 0$.  By a \emph{type} (term, small proposition, proposition, context, judgement) \emph{of $\mathcal{B}_n$} , we mean a type (term, small proposition, proposition, context, judgement) of $\LTTO$ that does not contain, as a subexpression, any type $A$ such that $D(A) > n$.

We say a judgement $\mathcal{J}$ of $\mathcal{B}_n$ is \emph{derivable} in $\mathcal{B}_n$ iff there exists a derivation of $\mathcal{J}$  in $\LTTO$ consisting
solely of judgements of $\mathcal{B}_n$; that is, a derivation of $\mathcal{J}$ in which no type $A$ occurs such that $D(A) > n$.  In this section, we write
$\Gamma \vdash_n \mathcal{J}$ iff the judgement $\Gamma \vdash \mathcal{J}$ is derivable in $\mathcal{B}_n$.
\end{df}

We define an interpretation of $\mathcal{B}_{n+1}$ in terms of $\mathcal{B}_n$:
\begin{df}
\label{df:interpB}
Fix $n \geq 1$.  Let $\Delta$ be a context of $\mathcal{B}_n$, and $v$ a function.  We define the following simultaneously.
\begin{itemize}
\item Given a term $M$ of $\mathcal{B}_{n+1}$, define the object $\round{B}^v$.
\begin{eqnarray*}
 \round{x}^v & \simeq & v(x) \\
\round{0}^v & \equiv & 0 \\
\round{\s M}^v & \simeq & \s \round{M}^v \\
\lefteqn{\round{\EN([x]T(K), L, [x,y]M, N)}^v} \\ & \simeq & \EN([x]T(\round{K}^{v[x:=x]}), \round{L}^v, \\
& & \quad [x,y]\round{M}^{v[x:=x,y:=y]}, \round{N}^v) \\
\round{(M,N)_{A \times B}}^v & \simeq & \begin{cases}
                            (\round{M}^v, \round{N}^v)_{\round{A}^v \times \round{B}^v} & \mbox{if } D(A \times B) \leq  n \\
\langle \round{M}^v, \round{N}^v \rangle & \mbox{if } D(A \times B) =  n + 1
                           \end{cases} \\
\round{\pi_1^{A \times B}(M)}^v & \simeq & \begin{cases}
                               \pi_1^{\round{A}^v \times \round{B}^v}(\round{M}^v) & \mbox{if } D(A \times B) \leq  n \\
\Pi_1(\round{M}^v) & \mbox{if } D(A \times B) =  n + 1
                              \end{cases} \\
\round{\pi_2^{A \times B}(M)}^v & \simeq & \begin{cases}
                               \pi_2^{\round{A}^v \times \round{B}^v}(\round{M}^v) & \mbox{if } D(A \times B) \leq  n \\
\Pi_2(\round{M}^v) & \mbox{if } D(A \times B) =  n + 1
                              \end{cases} \\
\round{\lambda x:A.M:B}^v & \simeq & \begin{cases}
                                    \lefteqn{\lambda x : \round{A}^v. \round{M}^{v[x:=x]} : \round{B}^v} & \\
 & \mbox{if } D(A \rightarrow B) \leq  n \\
\langle x, \round{M}^{v[x:=x]} \rangle 
 & \mbox{if } D(A \rightarrow B) =  n + 1
                                   \end{cases} \\
\round{M(N)_{A \rightarrow B}}^v & \simeq & \begin{cases}
                           \round{M}^v(\round{N}^v)_{\round{A}^v \rightarrow \round{B}^v} & \mbox{if } D(A \rightarrow B) \leq  n \\
\round{M}^v @ \round{N}^v & \mbox{if } D(A \rightarrow B) =  n + 1
                          \end{cases} \\
\round{\hat{N}}^v & \equiv & \hat{\N} \\
\round{M \hat{\times} N}^v & \simeq & \round{M}^v \hat{\times} \round{N}^v \\
\round{\{ x : A \mid P \}}^v & \simeq & \begin{cases}
                                         \{ x : \round{A}^v \mid \round{P}^{v[x:=x]} \} & \mbox{if } D(\Set{A}) \leq  n \\
\langle x, \round{P}^{v[x:=x]} \rangle & \mbox{if } D(\Set{A}) =  n + 1
                                        \end{cases}
\addtocounter{equation}{1} \end{eqnarray*}
\item Given a type $A$ of $\mathcal{B}_{n+1}$ such that $D(A) \leq  n$, define the type $\round{A}^v$ of $\mathcal{B}_n$.
\begin{eqnarray*}
\round{\N}^v & \equiv & \N \\
\round{A \times B}^v & \simeq & \round{A}^v \times \round{B}^v \\
\round{A \rightarrow B}^v & \simeq & \round{A}^v \rightarrow \round{B}^v \\
\round{U}^v & \equiv & U \\
\round{T(M)}^v & \simeq & T(\round{M}^v) \\
\round{\Set{A}}^v & \simeq & \Set{\round{A}^v}
\addtocounter{equation}{1} \end{eqnarray*}
\item
Given a small proposition $P$ of $\mathcal{B}_{n+1}$, define the small proposition $\round{P}^v$ as follows.
\begin{eqnarray*}
 \round{M_1 \hat{=}_N M_2}^v & \simeq & \round{M_1}^v \hat{=}_{\round{N}^v} \round{M_2}^v \\
\round{\hat{\bot}}^v & \equiv & \hat{\bot} \\
\round{P \hat{\supset} Q}^v & \simeq & \round{P}^v \hat{\supset} \round{Q}^v \\
\round{\hat{\forall} x : M. P}^v & \simeq & \hat{\forall} x : \round{M}^v . \round{P}^{v[x:=x]} \\
\round{M \hat{\in}_A N}^v & \simeq & \round{M}^v \hat{\in}_{\round{A}^v} \round{N}^v
\addtocounter{equation}{1} \end{eqnarray*}
\item Given a type $A$ of $\mathcal{B}_{n+1}$, define a set $\brackets{A}_\Delta^v$ and an equivalence relation $\sim_{\Delta v}^A$ on this set.

If $D(A) \leq  n$, then
\begin{eqnarray*}
 \brackets{A}_\Delta^v & = & \{ M \mid \Delta \vdash_n M : \round{A}^v \} \\
M \sim_{\Delta v}^A N & \Leftrightarrow & \Delta \vdash_n M = N : \round{A}^v
\addtocounter{equation}{1} \end{eqnarray*}
Otherwise,
\begin{eqnarray*}
 \brackets{A \times B}_\Delta^v & = & \{ \langle M,N \rangle \mid \Delta \vdash_n M : \round{A}^v , \Delta \vdash_n N : \round{B}^v \} \\
\langle M,N \rangle \sim_{\Delta v}^{A \times B} \langle M', N' \rangle & \Leftrightarrow & \Delta \vdash_n M = M' : \round{A}^v \\
& & \quad \wedge \Delta \vdash_n N = N' : \round{B}^v \\
& & \\
\brackets{A \rightarrow B}_\Delta^v & = & \{ \langle x,M \rangle \mid \Delta, x : \round{A}^v \vdash M : \round{B}^v \} \\
\langle x,M \rangle \sim_{\Delta v}^{A \rightarrow B} \langle x, M' \rangle & \Leftrightarrow & \Delta, x : \round{A}^v \vdash M = M' : \round{B}^v \\
& & \\
\brackets{\Set{A}}_\Delta^v & = & \{ \langle x,P \rangle \mid \Delta, x : \round{A}^v \vdash P \p \} \\
\langle x,P \rangle \sim_{\Delta v}^{\Set{A}} \langle x, P' \rangle & \Leftrightarrow & \Delta, x : \round{A}^v \vdash P = P'
\addtocounter{equation}{1} \end{eqnarray*}
\item
Given a context $\Gamma \equiv x_1 : A_1, \ldots, x_m : A_m$ of $\mathcal{B}_{n+1}$, we say that $v$ is a \emph{$\Delta$-valuation} of $\Gamma$ iff $v(x_i) \in \brackets{A_i}_\Delta^v$ for each $i$.
\item
Given a proposition $\phi$ of $\mathcal{B}_{n+1}$ such that $D(\phi) \leq  n$, define the proposition $\round{\phi}^v$ as follows.
\begin{eqnarray*}
 \round{M_1 =_N M_2}^v & \simeq & \round{M_1}^v =_{\round{N}^v} \round{M_2}^v \\
\round{\bot}^v & \equiv & \bot \\
\round{\phi \supset \psi}^v & \simeq & \round{\phi}^v \supset \round{\psi}^v \\
\round{\forall x : A. \phi}^v & \simeq & \forall x : \round{A}^v. \round{\phi}^{v[x:=x]} \\
\round{V(P)}^v & \simeq & V(\round{P}^v)
\addtocounter{equation}{1} \end{eqnarray*}
\end{itemize}
\end{df}

We define what the notion of \emph{satisfaction} $(\Delta, \Phi, v) \models \psi$ similarly to Definition \ref{df:satis}:
\begin{df}[Satisfaction]
Let $\Phi \equiv \phi_1, \ldots, \phi_m$ be a sequence of propositions of $\mathcal{A}_n$
such that $\Delta \vdash_n \phi_1 \prop$, \ldots, $\Delta \vdash_n \phi_m \prop$.  Let $v$ be a $\Delta$-valuation of $\Gamma$.  Suppose $\Gamma \vdash_{n+1} \phi \prop$.  We define what it means for $(\Delta, \Phi, v)$ to \emph{satisfy} $\phi$, $(\Delta, \Phi, v) \models \phi$, as follows.

If $D(\phi) \leq n$, then
$((\Delta, \Phi, v) \models \phi) \Leftrightarrow (\Delta \vdash_n \Phi \Rightarrow \round{\phi}^v)$.

Otherwise,
\begin{itemize}
\item
$(\Delta, \Phi, v) \models \phi \supset \psi$ iff, for all $\Delta' \supseteq \Delta$ and $\Phi' \supseteq \Phi$, if $(\Delta', \Phi', v) \models \phi$ then $(\Delta', \Phi', v) \models \psi$.
\item
$(\Delta, \Phi, v) \models \forall x:A.\phi$ iff, for all $\Delta' \supseteq \Delta$ and $a \in \brackets{A}_{\Delta'}^v$, we have \\
$(\Delta', \Phi, v[x:=a]) \models \phi$.
\end{itemize}
\end{df}

\begin{df}[Satisfaction and Truth]
Let $\Gamma \vdash \mathcal{J}$ be a judgement of $\mathcal{B}_{n+1}$.
Let $\Delta \vdash_n \vald$ and $v$ be a $\Delta$-valuation of $\Gamma$.  We define what it means for $\Delta$ and $v$ to \emph{satisfy} $\mathcal{J}$, $(\Delta, v) \models \mathcal{J}$, as follows.
\begin{itemize}
\item
If $D(A) \leq n$, then $(\Delta, v) \models A \type$ iff $\brackets{A}_\Delta^v$ is defined and $\Delta \vdash_n \round{A}^v \type$.

If $D(A) = n+1$, then $(\Delta, v) \models A \type$ iff $\brackets{A}_\Delta^v$ is defined.
\item
If $D(A), D(B) \leq n$, then $(\Delta, v) \models A = B$ iff $\brackets{A}_\Delta^v = \brackets{B}_\Delta^v$ and \\ $(\sim_{\Delta v}^A) = (\sim_{\Delta v}^B)$ and $\Delta \vdash_n \round{A}^v = \round{B}^v$.

If $D(A) = D(B) = n+ 1$, then $(\Delta, v) \models A = B$ iff $\brackets{A}_\Delta^v = \brackets{B}_\Delta^v$ and $(\sim_{\Delta v}^A) = (\sim_{\Delta v}^B)$.
\item
$(\Delta, v) \models M : A$ iff $\round{M}^v \in \brackets{A}_\Delta^v$.
\item
$(\Delta, v) \models M = N : A$ iff $\round{M}^v \sim_{\Delta v}^A \round{N}^v$
\item
$(\Delta, v) \models P \p$ iff $\Delta \vdash_n \round{P}^v \p$.
\item
$(\Delta, v) \models P = Q$ iff $\Delta \vdash_n \round{P}^v = \round{Q}^v$
\item
If $D(\phi) \leq n$, then $(\Delta, v) \models \phi \prop$ iff $\Delta \vdash_n \round{\phi}^v \prop$.
\item
$(\Delta, v) \models \phi = \psi$ iff, for all $\Phi$, we have $(\Delta, \Phi, v) \models \phi$ iff $(\Delta, \Phi, v) \models \psi$.
\item
$(\Delta, v) \models \psi_1, \ldots, \psi_m \Rightarrow \chi$ iff, for all $\Phi$, if $(\Delta, \Phi, v)$ satisfies $\psi_i$ for all $i$, then $(\Delta, \Phi, v)$ satisfies $\chi$.
\item
For any other $\mathcal{J}$, we have $(\Delta, v) \models \mathcal{J}$ for all $\Delta$, $v$.
\end{itemize}
We say $\Gamma \vdash \mathcal{J}$ is \emph{true} iff, whenever $\Delta \vdash_n \vald$ and $v$ is a $\Delta$-valuation of $\Gamma$, then $(\Delta, v) \models \mathcal{J}$.
\end{df}

\begin{thm}[Soundness]
Every derivable judgement in $\mathcal{B}_{n+1}$ is true.
\end{thm}

\begin{pf}
Similar to Theorems \ref{thm:soundness} and \ref{thm:soundness'}.
\end{pf}

\begin{thm}$ $
\begin{enumerate}
\item
Let $\Gamma \vdash \mathcal{J}$ be a judgement of $\mathcal{B}_n$, and suppose $\mathcal{J}$ does not have the form $\Phi \Rightarrow \psi$.
If the judgement is true, and $\Gamma \vdash_n \vald$, then the judgement is derivable in $\mathcal{B}_n$.
\item
Let $\Gamma \vdash \phi_1, \ldots, \phi_m \Rightarrow \psi$ be a judgement of $\mathcal{B}_n$.  If the judgement is true, and we have $\Gamma \vdash_n \vald$ and $\Gamma \vdash_n \phi_i \prop$ for $i=1, \ldots, m$, then the judgement is derivable in $\mathcal{B}_n$.
\end{enumerate}
\end{thm}

\begin{pf}
Similar to Theorem \ref{thm:completeness}.
\end{pf}

\begin{cor}
 If $\mathcal{J}$ is a judgement of $\mathcal{B}_n$ derivable in $\mathcal{B}_{n+1}$, then $\mathcal{J}$ is derivable in $\mathcal{B}_n$.
\end{cor}

\begin{cor}
\label{cor:LTTOTwU}
 If $\mathcal{J}$ is a judgement of $\TwU$ derivable in $\LTTO$, then $\mathcal{J}$ is derivable in $\TwU$.
\end{cor}

With this final step, we have now completed the proof of the conservativity of $\LTTO$ over $\ACAO$:

\begin{cor}
Let $\phi$ be a formula of second order arithmetic with free variables $x_1$, \ldots, $x_m$, $X_1$, \ldots, $X_n$.  If
\[ x_1 : \N, \ldots, x_m : \N, X_1 : \Set{\N}, \ldots, X_n : \Set{\N} \vdash \Rightarrow \angles{\phi} \]
in $\LTTO$ then $\ACAO \vdash \phi$.
\end{cor}

\begin{pf}
Let $\mathcal{J}$ be the judgement \\
$x_1 : \N, \ldots, x_m : \N, X_1 : \Set{\N}, \ldots, X_n : \Set{\N} \vdash \Rightarrow \angles{\phi}$.

Suppose $\mathcal{J}$ is derivable in $\LTTO$.  Then
\[ \begin{array}{cl}
\mathcal{J} \mbox{ is derivable in } \TwU & (\mbox{Corollary \ref{cor:LTTOTwU}}) \\
\therefore \mathcal{J} \mbox{ is derivable in } \Tw & (\mbox{Corollary \ref{cor:TwUTw}}) \\
\therefore \mathcal{J} \mbox{ is derivable in } \Tt & (\mbox{Corollary \ref{cor:TwTt}}) \\
\therefore \ACAO \vdash \phi & (\mbox{Corollary \ref{cor:TtACAO}})
\end{array} \]
\end{pf}

\section{Other Conservativity Results}

\subsection{Conservativity of $\LTTOs$ over ACA}

Our proof method can be adapted quite straightforwardly to prove the conservativity of $\LTTOs$ over ACA.  We shall present these proofs briefly, giving only the details that need to be changed.

We define subsystems of $\LTTOs$:
\[ \Tt^* \hookrightarrow \Tw^* \hookrightarrow \TwU^* \hookrightarrow \LTTOs \]
$\Tt^*$ is formed from $\Tt$ by allowing the rule $(\IndN)$ to be applied with any analytic proposition $\phi$.  In the same manner, $\Tw^*$ is formed from $\Tw$, $\TwU^*$ is formed from $\TwU$, and $\LTTOs$ is formed from $\LTTO$.

The proof of the conservativity of $\LTTOs$ over $\Tt^*$ follows exactly the same pattern as in Section \ref{section:conservativity}.

\begin{thm}
 Theorem \ref{thm:soundphi} holds for $\Tt^*$ and $\ACA$.
\end{thm}

\begin{pf}
Similar to the proof of Theorem \ref{thm:soundphi}.
\end{pf}

Similarly, Corollary \ref{cor:TwTt} holds for $\Tw^*$ and $\Tt^*$, Corollary \ref{cor:TwUTw} holds for $\TwU^*$ and $\Tw^*$, and Corollary \ref{cor:LTTOTwU} holds for $\LTTOs$ and $\TwU^*$.  This completes the proof that $\LTTOs$ is conservative over $\ACA$.

\subsection{Conservativity of $\ACAO$ over PA}
\label{section:ACAOPA}
As a side-benefit of this work, we can easily produce as a corollary another proof that $\ACAO$ is conservative over Peano Arithmetic (PA).  We can define a system $T_1$ with just one type, $\N$, in its type-theoretic component.  We can apply our method to show that $T_2$ is conservative over $T_1$, and that $T_1$ is conservative over PA; we omit the details.  

Combining all these proofs, we can produce the following elementary proof that $\ACAO$ is conservative over PA, which proceeds by interpreting the formulas of $\ACAO$ as statements about PA. To the best of the authors' knowledge, this proof has not appeared in print before.
\begin{thm}
$\ACAO$ is conservative over PA.
\end{thm}

\begin{pf}
Define a \emph{PA-formula} to be a formula in which no set variables (bound or free) occur.

Let $\mathcal{V}$ be a set of variables of $\L$.  A \emph{valuation} of $\mathcal{V}$ is a function $v$ on $\mathcal{V}$ such that:
\begin{itemize}
\item
for every number variable $x \in \mathcal{V}$, $v(x)$ is a term of PA;
\item
for every set variable $X \in \mathcal{V}$, $v(X)$ is an expression of the form $\{ y \mid \phi \}$ where $\phi$ is a PA-formula.
\end{itemize}
For $t$ a term, let $v(t)$ be the result of substituting $v(x)$ for each variable $x$ in $t$.

For $\phi$ a formula of $\L$, let $v(\phi)$ be the PA-formula that results from making the following replacements throughout $\phi$.
\begin{itemize}
\item
Replace each atomic formula $s = t$ with $v(s) = v(t)$.
\item
For each atomic formula $t \in X$, let $v(X) = \{ y \mid \psi \}$.  Replace $t \in X$ with $[v(t)/y]\psi$.
\end{itemize}

Define what it is for a valuation $v$ and PA-formula $\psi$ to \emph{satisfy} an $\L$-formula $\phi$, $(v,\psi) \models \phi$, as follows.
\begin{itemize}
\item
If $\phi$ is arithmetic, $(v,\psi) \models \phi$ iff $\psi \supset v(\phi)$ is a theorem of PA.  Otherwise:
\item
$(v,\psi) \models \phi \supset \chi$ iff, for any PA-formula $\psi'$, if $(v, \psi \wedge \psi') \models \phi$ \\ then $(v, \psi \wedge \psi') \models \chi$.
\item
$(v, \psi) \models \forall x \phi$ iff, for every term $t$, $(v[x:=t], \psi) \models \phi$.
\item
$(v, \psi) \models \forall X \phi$ iff, for every PA-formula $\chi$, $(v[X:=\{y \mid \chi\}], \psi) \models \phi$.
\end{itemize}
Let us say that a formula $\phi$ of $\L$ is \emph{true} iff $(v, x = x) \models \phi$ for every valuation $v$.

We prove the following two claims:
\begin{enumerate}
\item
Every theorem of $\ACAO$ is true.
\item
Every PA-formula that is true is a theorem of PA.
\end{enumerate}
The first claim is proven by induction on derivations in $\ACAO$.  As an example, consider the axiom
\[ \forall X (\phi \supset \psi) \supset (\phi \supset \forall X \psi) \]
where $X \notin \FV{\phi}$.  Fix $v$ and $\chi$, and suppose
\[ (v, \chi) \models \forall X (\phi \supset \psi) \enspace . \]
We must show that $(v, \chi) \models \phi \supset \forall X \psi$.

Let $\chi'$ be any PA-formula, and suppose $(v, \chi \wedge \chi') \models \phi$.  Let $\tau$ be any PA-formula; we must show that
$(v[X := \{ y \mid \tau \}], \chi \wedge \chi') \models \phi$.  Since $X \notin \FV{\phi}$, we have that
\[ (v[X := \{ y \mid \tau \}], \chi \wedge \chi') \models \phi \]
We also have $(v[X := \{ y \mid \tau \}], \chi \wedge \chi') \models \phi \supset \psi$, and so $(v[X := \{ y \mid \tau \}], \chi \wedge \chi') \models \psi$
as required.

The second claim is proven using the valuation that is the identity on $\FV{\phi}$.

It follows that, if a formula of PA is a theorem of $\ACAO$, then it is a theorem of PA.
\end{pf}

\paragraph{Remarks}
\begin{enumerate}
\item
The same method could be used to show that G\"odel-Bernays set theory is conservative over ZF set theory.
\item
Another proof-theoretic method of proving this results is given in Shoenfield \cite{shoenfield:rcp}.  That proof relies on some quite strong results about classical theories; our proof is more elementary.  However, Shoenfield's proof is constructive (giving an algorithm that would produce a proof of $\bot$ in PA from a proof of $\bot$ in $\ACAO$) and can be formalised in PRA; ours has neither of these properties.
\end{enumerate}

\subsection{$\ACAOp$}
An argument has been made that the system $\ACAOp$ corresponds to Weyl's foundation \cite[p.135]{bs:hb}, claiming that its axiom schema of $\omega$-iterated arithmetical comprehension `occurs in the formal systems defined by Weyl and Zahn', presumably a reference to Weyl's \emph{Principle of Iteration} \cite[p.38]{weyl:continuum}.

The axioms of $\ACAOp$ are the axioms of $\ACAO$ together with the following \emph{axiom schema of $\omega$-iterated arithmetical comprehension}.  Assume we have defined a pairing function $(x,y)$ in $\ACAO$.  We put
\[ (X)_j = \{ n : (n,j) \in X \}, \qquad (X)^j = \{ (m,i) : (m,i) \in X \wedge i < j \} \enspace . \]
Then, for every arithmetical formula $\phi[n,Y]$ in which $X$ does not occur free, the following is an axiom:
\[ \exists X \forall j \forall n(n \in (X)_j \leftrightarrow \phi[n, (X)^j]) \enspace . \]

The translation we gave in Section \ref{section:embed} is a sound translation from $\ACAOp$ into $\LTTW$.  It is difficult to construct a subsystem of $\LTTW$ that is conservative over $\ACAOp$, however.  A natural suggestion would be to extend $\LTTO$ by allowing $\EN$ to take either a small type, or the type $\Set{\N}$; let us call the system produced $\LTTO^+$.  Then $\LTTO^+$ is indeed conservative over $\Tt^+$, the extension of $\Tt$ with a new constructor
\[ \begin{prooftree}
\begin{array}{c}
\Gamma \vdash L : \Set{\N}
\quad
\Gamma, x : \N, Y : \Set{\N} \vdash M : \Set{\N} \\
\Gamma \vdash N : \N
\end{array}
\justifies
\Gamma \vdash \R^+(L, [x,Y]M, N) : \Set{\N}
   \end{prooftree} \]
and appropriate equality rules.

However, it seems unlikely that $\Tt^+$ is conservative over $\ACAOp$.  In particular, there seems to be no way to interpret terms that involve \emph{two} or more applications of $R^+$.  In $\LTTO^+$, we may iterate \emph{any} definable function $\Set{\N} \rightarrow \Set{\N}$.  In $\ACAOp$, we may only iterate those functions that are defined by an arithmetic proposition; and not every such function definable in $\ACAOp$ is defined by an arithmetic proposition.

\section{Conclusion}

We have constructed two subsystems of  $\LTTW$, and proved that these are conservative over $\ACAO$ and $\ACA$ respectively.  We have thus shown how, using LTTs, we can take a system like $\ACAO$ or $\ACA$ and add to it the ability to speak of pairs, functions of all orders, sets of all orders, and a universe of types, without increasing the proof-theoretic strength of the system.

We have also begun the proof-theoretic analysis of $\LTTW$.
We now know that $\LTTW$ is strictly stronger than $\LTTO$, and hence $\ACAO$.  The subsystem $\LTTOs$ is quite a
small fragment of $\LTTW$, and so we conjecture that $\LTTW$ is strictly stronger than $\LTTOs$, and hence strictly stronger than $\ACA$.  Once this conjecture is proven, we will have quite strong evidence for our claim that Weyl's foundation exceeds both $\ACAO$ and $\ACA$.

The method of proof we have given is quite a general one, and should be applicable in many other situations.  It does not rely on any reduction properties of the type system, and so could be applied to type systems that are not strongly normalising, or do not satisfy Church-Rosser (or are not known to be strongly normalising or to satisfy Church-Rosser).
It provides a uniform method for proving types redundant; we were able to remove products, function types, types of sets, and the universe from $\LTTO$.

Furthermore, the method allowed us to separate these tasks.  We were able to remove $U$ separately from the other types, and to use a different interpretation to do so.  In Sections \ref{section:TwiTti} and \ref{section:LTTOTwU}, for example, we interpreted judgemental equality by judgemental equality; in Section \ref{section:TwUTw}, we interpreted judgemental equality by propositional equality.
Our method is thus quite powerful; we did not have to find a single interpretation that would perform all these tasks.

A proof of our conjecture that $\LTTW$ is stronger than $\LTTOs$ has very recently been discovered, by the first author and Anton Setzer.
The proof theoretic strength of $\LTTW$ is in fact $\phi_{\epsilon_0}(0)$.  A paper presenting the proof of this result is in
preparation.

For future work, we should investigate more generally how adding features to an LTT changes its proof-theoretic strength.  This will be a more
difficult task, as we will need to investigate what effect induction and recursion have when they are no longer confined to the small types and propositions.
We are particularly interested in the differences between LTTs and systems of predicate logic; for example, in how the strength of an LTT changes when we modify the type-theoretic component but not the logical component.

Finally, we note that there are striking superficial similarities between our work and Streicher \cite{streicher:stt}, who also gave interpretations to type theories.  Like our interpretations, his were first defined as partial functions on the syntax, then proven to be total on the typable terms by induction on derivations.  He also made use of a `depth' function on types.  Our work is not a direct application of his, but it remains to be seen whether there are formal connections that can be exploited.

\appendix

\section{Formal Definition of Systems}

We present here the definition of $\LTTW$ and the two principal subsystems used in this paper.

\subsection{$\LTTW$}
\label{app:lttw}
The syntax of $\LTTW$ is given by the following grammar:
\[ \begin{array}{lrcl}
\mbox{Type} & A & ::= & \N \mid A \times A \mid A \rightarrow A \mid U \mid T(M) \mid \Set{A} \\
\mbox{Term} & M & ::= & x \mid 0 \mid \s M \mid \EN([x]A, M, [x,x]M, M) \mid \\
& & & (M,M)_{A \times A} \mid \pi_1^{A \times A}(M) \mid \pi_2^{A \times A}(M) \mid \\
& & & \lambda x:A.M:A \mid M(M)_{A \rightarrow A} \mid \hat{\N} \mid M \hat{\times} M \mid \\
& & & \{ x : A \mid P \} \\
\mbox{small Proposition} & P & ::= & M \hat{=}_M M \mid \hat{\bot} \mid P \hat{\supset} P \mid \hat{\forall} x : M. P \mid M \hat{\in}_A M \\
\mbox{Formula} & \phi & ::= & M =_M M \mid \bot \mid \phi \supset \phi \mid \forall x:A.\phi \mid V(P)
   \end{array} \]

We write $\neg \phi$ for $\phi \supset \bot$, and $M \in_A N$ for $V(M \hat{\in}_A N)$.

The rules of deduction of $\LTTW$ are as follows:

\subsubsection{Structural Rules}
\label{section:LTTs}
\[ \begin{prooftree}
    \justifies
 \vdash \vald
   \end{prooftree}
\qquad
\begin{prooftree}
 \Gamma \vdash A \type
\justifies
\Gamma, x : A \vdash \vald
\end{prooftree}
\qquad
\begin{prooftree}
 \Gamma \vdash \vald
\using{(x : A \in \Gamma)}
\justifies
\Gamma \vdash x : A
\end{prooftree} \]

\[ \begin{prooftree}
    \Gamma \vdash M : A
\justifies
\Gamma \vdash M = M : A
   \end{prooftree}
\qquad
\begin{prooftree}
 \Gamma \vdash M = N : A
\justifies
\Gamma \vdash N = M : A
\end{prooftree}
\qquad
\begin{prooftree}
 \Gamma \vdash M = N : A
\quad
\Gamma \vdash N = P : A
\justifies
\Gamma \vdash M = P : A
\end{prooftree} \]

\[ \begin{prooftree}
    \Gamma \vdash A \type
\justifies
\Gamma \vdash A = A
   \end{prooftree}
\qquad
\begin{prooftree}
 \Gamma \vdash A = B
\justifies
\Gamma \vdash B = A
\end{prooftree}
\qquad
\begin{prooftree}
 \Gamma \vdash A = B
\quad
\Gamma \vdash B = C
\justifies
\Gamma \vdash A = C
\end{prooftree} \]

\[ \begin{prooftree}
    \Gamma \vdash M : A
\quad
\Gamma \vdash A = B
\justifies
\Gamma \vdash M : B
   \end{prooftree}
\qquad
\begin{prooftree}
 \Gamma \vdash M = N : A
\quad
\Gamma \vdash A = B
\justifies
\Gamma \vdash M = N : B
\end{prooftree} \]

\[ \begin{prooftree}
    \Gamma \vdash P \p
\justifies
\Gamma \vdash P = P
   \end{prooftree}
\qquad
\begin{prooftree}
 \Gamma \vdash P = Q
\justifies
\Gamma \vdash Q = P
\end{prooftree}
\qquad
\begin{prooftree}
 \Gamma \vdash P = Q
\quad
\Gamma \vdash Q = R
\justifies
\Gamma \vdash P = R
\end{prooftree} \]

\[ \begin{prooftree}
    \Gamma \vdash \phi \prop
\justifies
\Gamma \vdash \phi = \phi
   \end{prooftree}
\qquad
\begin{prooftree}
 \Gamma \vdash \phi = \psi
\justifies
\Gamma \vdash \psi = \phi
\end{prooftree}
\qquad
\begin{prooftree}
 \Gamma \vdash \phi = \psi
\quad
\Gamma \vdash \psi = \chi
\justifies
\Gamma \vdash \phi = \chi
\end{prooftree} \]

\[ \begin{prooftree}
    \Gamma \vdash \phi_1 \prop
\quad
\cdots
\quad
\Gamma \vdash \phi_n \prop
\justifies
\Gamma \vdash \phi_1, \ldots, \phi_n \Rightarrow \phi_i
   \end{prooftree}
\quad
\begin{prooftree}
 \Gamma \vdash \Phi \Rightarrow \phi
\quad
\Gamma \vdash \phi = \psi
\justifies
\Gamma \vdash \Phi \Rightarrow \psi
\end{prooftree} \]

\subsubsection{Natural Numbers}

\[ \begin{prooftree}
    \Gamma \vdash \vald
\justifies
\Gamma \vdash \N \type
   \end{prooftree}
\qquad
\begin{prooftree}
 \Gamma \vdash \vald
\justifies
\Gamma \vdash 0 : \N
\end{prooftree}
\qquad
\begin{prooftree}
 \Gamma \vdash M : \N
\justifies
\Gamma \vdash \s M : \N
\end{prooftree}
\qquad
\begin{prooftree}
 \Gamma \vdash M = M' : \N
\justifies
\Gamma \vdash \s M = \s M' : \N
\end{prooftree} \]

\[ (\EN) \; \begin{prooftree}
\begin{array}{cc}
    \Gamma, x : \N \vdash C \type
&
\Gamma \vdash L : [0/x]C \\
\Gamma, x : \N, y : C \vdash M : [\s x / x]C
&
\Gamma \vdash N : \N
\end{array}
\justifies
\Gamma \vdash \EN([x]C, L, [x,y]M, N) : [N/x]C
   \end{prooftree} \]

\[ (\EN=) \; \begin{prooftree}
\begin{array}{cc}
    \Gamma, x : \N \vdash C = C'
&
\Gamma \vdash L = L' : [0/x]C \\
\Gamma, x : \N, y : C \vdash M = M' : [\s x / x]C
&
\Gamma \vdash N = N' : \N
\end{array}
\justifies
\Gamma \vdash \EN([x]C, L, [x,y]M, N) = \EN([x]C', L', [x,y]M', N') : [N/x]C
   \end{prooftree} \]

\[ (\EN0) \; \begin{prooftree}
\begin{array}{c}
 \Gamma, x : \N \vdash C \type
\quad
\Gamma \vdash L : [0/x]C \\
\Gamma, x : \N, y : C \vdash M : [\s x / x] C
\end{array}
\justifies
\Gamma \vdash \EN([x]C, L, [x,y]M, 0) = L : [0/x]C
   \end{prooftree} \]

\[ (\EN\s) \; \begin{prooftree}
\begin{array}{cc}
    \Gamma, x : \N \vdash C \type
&
\Gamma \vdash L : [0/x]C \\
\Gamma, x : \N, y : C \vdash M : [\s x / x]C
&
\Gamma \vdash N : \N
\end{array}
\justifies
\begin{array}{l}
\Gamma \vdash \EN([x]C, L, [x,y]M, \s N) \\
= [N/x, \EN([x]C, L, [x,y]M, N) / y] M : [\s N/x]C
\end{array}
   \end{prooftree} \]

\[ (\IndN) \begin{prooftree}
\begin{array}{cc}
            \Gamma, x : \N \vdash \phi \prop
& \Gamma \vdash N : \N \\
\Gamma \vdash \Phi \Rightarrow [0/x]\phi
& \Gamma, x : \N \vdash \Phi, \phi \Rightarrow [\s x/x]\phi
\end{array}
\justifies
\Gamma \vdash \Phi \Rightarrow [N/x]\phi
           \end{prooftree} \]

\subsubsection{Pairs}
\label{section:pairs}

\[ \begin{prooftree}
    \Gamma \vdash A \type
\quad
\Gamma \vdash B \type
\justifies
\Gamma \vdash A \times B \type
   \end{prooftree}
\qquad
\begin{prooftree}
 \Gamma \vdash A = A'
\quad
\Gamma \vdash B = B'
\justifies
\Gamma \vdash (A \times B) = (A' \times B')
\end{prooftree} \]

\[ \begin{prooftree}
    \Gamma \vdash M : A
\quad
\Gamma \vdash N : B
\justifies
\Gamma \vdash (M,N)_{A \times B} : A \times B
   \end{prooftree}
\qquad
\begin{prooftree}
\begin{array}{cc}
 \Gamma \vdash A = A' &
\Gamma \vdash B = B' \\
\Gamma \vdash M = M' : A &
\Gamma \vdash N = N' : B
\end{array}
\justifies
\Gamma \vdash (M,N)_{A \times B} = (M',N')_{A' \times B'} : A \times B
\end{prooftree} \]

\[ \begin{prooftree}
    \Gamma \vdash M : A \times B
\justifies
\Gamma \vdash \pi_1^{A \times B}(M) : A
   \end{prooftree}
\qquad
\begin{prooftree}
\begin{array}{c}
 \Gamma \vdash A = A' \quad \Gamma \vdash B = B' \\
 \Gamma \vdash M = M' : A \times B
\end{array}
\justifies
\Gamma \vdash \pi_1^{A \times B}(M) = \pi_1^{A' \times B'}(M') : A
\end{prooftree} \]

\[ \begin{prooftree}
    \Gamma \vdash M : A \times B
\justifies
\Gamma \vdash \pi_2^{A \times B}(M) : B
   \end{prooftree}
\qquad
\begin{prooftree}
\begin{array}{c}
 \Gamma \vdash A = A' \quad \Gamma \vdash B = B' \\
 \Gamma \vdash M = M' : A \times B
\end{array}
\justifies
\Gamma \vdash \pi_2^{A \times B}(M) = \pi_2^{A' \times B'}(M') : B
\end{prooftree} \]

\[ \begin{prooftree}
    \Gamma \vdash M : A
\quad
\Gamma \vdash N : B
\justifies
\Gamma \vdash \pi_1^{A \times B}((M,N)_{A \times B}) = M : A
   \end{prooftree}
\qquad
\begin{prooftree}
    \Gamma \vdash M : A
\quad
\Gamma \vdash N : B
\justifies
\Gamma \vdash \pi_2^{A \times B}((M,N)_{A \times B}) = N : B
   \end{prooftree} \]

\[ (\etaX) \begin{prooftree}
\begin{array}{c}
            \Gamma, z : A \times B \vdash \phi \prop
\quad
\Gamma \vdash M : A \times B \\
\Gamma \vdash \Phi \Rightarrow [(\pi_1^{A \times B}(M), \pi_2^{A \times B}(M)) / z] \phi
\end{array}
\justifies
\Gamma \vdash \Phi \Rightarrow [M/z] \phi
           \end{prooftree} \]

\subsubsection{Functions}
\label{section:functions}
\[ \begin{prooftree}
    \Gamma \vdash A \type
\quad
\Gamma \vdash B \type
\justifies
\Gamma \vdash A \rightarrow B \type
   \end{prooftree}
\qquad
\begin{prooftree}
 \Gamma \vdash A = A'
\quad
\Gamma \vdash B = B'
\justifies
\Gamma \vdash (A \rightarrow B) = (A' \rightarrow B')
\end{prooftree} \]

\[ \begin{prooftree}
    \Gamma, x : A \vdash M : B
\justifies
\Gamma \vdash (\lambda x:A.M:B) : A \rightarrow B
   \end{prooftree}
\qquad
\begin{prooftree}
\begin{array}{c}
 \Gamma \vdash A = A'
\quad
\Gamma \vdash B = B' \\
\Gamma, x : A \vdash M = M' : B
\end{array}
\justifies
\begin{array}{l}
\Gamma \vdash (\lambda x:A.M:B) \\
\quad = (\lambda x:A'.M':B') : A \rightarrow B
\end{array}
\end{prooftree} \]

\[ \begin{prooftree}
    \Gamma \vdash M : A \rightarrow B
\quad
\Gamma \vdash N : A
\justifies
\Gamma \vdash M(N)_{A \rightarrow B} : B
   \end{prooftree}
\qquad
\begin{prooftree}
\begin{array}{cc}
 \Gamma \vdash A = A'
&
\Gamma \vdash B = B' \\
 \Gamma \vdash M = M' : A \rightarrow B
&
\Gamma \vdash N = N' : A
\end{array}
\justifies
\Gamma \vdash M(N)_{A \rightarrow B} = M'(N')_{A' \rightarrow B'} : B
\end{prooftree} \]

\[ \begin{prooftree}
    \Gamma, x: A \vdash M : B
\quad
\Gamma \vdash N : A
\justifies
\Gamma \vdash (\lambda x:A.M:B)(N)_{A \rightarrow B} = [N/x]M : [N/x]B
   \end{prooftree} \]

\[ (\etaA) \; \begin{prooftree}
\begin{array}{c}
               \Gamma, z: A \rightarrow B \vdash \phi \prop
\quad
\Gamma \vdash M : A \rightarrow B \\
\Gamma \vdash \Phi \Rightarrow [\lambda x:A.M(x):B / z]\phi
\end{array}
\justifies
\Gamma \vdash \Phi \Rightarrow [M/z]\phi
              \end{prooftree} \]

\subsubsection{Typed Sets}
\label{section:sets}
\[ \begin{prooftree}
    \Gamma \vdash A \type
\justifies
\Gamma \vdash \Set{A} \type
   \end{prooftree}
\qquad
\begin{prooftree}
 \Gamma \vdash A = A'
\justifies
\Gamma \vdash \Set{A} = \Set{A'}
\end{prooftree} \]

\[ \begin{prooftree}
    \Gamma, x : A \vdash P \p
\justifies
\Gamma \vdash \{ x : A \mid P \} : \Set{A}
   \end{prooftree}
\qquad
\begin{prooftree}
 \Gamma \vdash A = A'
\quad
\Gamma, x : A \vdash P = P'
\justifies
\Gamma \vdash \{ x : A \mid P \} = \{ x : A' \mid P' \} : \Set{A}
\end{prooftree} \]

\[ \begin{prooftree}
    \Gamma \vdash M : A
\quad
\Gamma \vdash N : \Set{A}
\justifies
\Gamma \vdash M \hat{\in}_A N \p
   \end{prooftree}
\quad
\begin{prooftree}
\begin{array}{c}
\Gamma \vdash A = A' \\
\Gamma \vdash M = M' : A
\quad
\Gamma \vdash N = N' : \Set{A}
\end{array}
\justifies
\Gamma \vdash (M \hat{\in}_A N) = (M' \hat{\in}_{A'} N')
\end{prooftree} \]

\[
\begin{prooftree}
 \Gamma \vdash M : A
\quad
\Gamma, x : A \vdash P \p
\justifies
\Gamma \vdash (M \hat{\in}_A \{ x : A \mid P \}) = [M/x]P
\end{prooftree} \]

\subsubsection{The Type Universe}

\[ \begin{prooftree}
    \Gamma \vdash \vald
\justifies
\Gamma \vdash U \type
   \end{prooftree}
\qquad
\begin{prooftree}
 \Gamma \vdash M : U
\justifies
\Gamma \vdash T(M) \type
\end{prooftree}
\qquad
\begin{prooftree}
 \Gamma \vdash M = M' : U
\justifies
\Gamma \vdash T(M) = T(M')
\end{prooftree} \]

\[ \begin{prooftree}
    \Gamma \vdash \vald
\justifies
\Gamma \vdash \hat{\N} : U
   \end{prooftree}
\qquad
\begin{prooftree}
 \Gamma \vdash \vald
\justifies
\Gamma \vdash T(\hat{\N}) = \N
\end{prooftree} \]

\[ \begin{prooftree}
    \Gamma \vdash M : U
\quad
\Gamma \vdash N : U
\justifies
\Gamma \vdash M \hat{\times} N : U
   \end{prooftree}
\qquad
\begin{prooftree}
 \Gamma \vdash M = M' : U
\quad
\Gamma \vdash N = N' : U
\justifies
\Gamma \vdash (M \hat{\times} M') = (N \hat{\times} N') : U
\end{prooftree} \]

\[ \begin{prooftree}
    \Gamma \vdash M : U
\quad
\Gamma \vdash N : U
\justifies
\Gamma \vdash T(M \hat{\times} N) = T(M) \times T(N)
   \end{prooftree} \]

\subsubsection{Classical Predicate Logic}
\label{section:predlog}
\[ \begin{prooftree}
\Gamma \vdash \vald
\justifies
\Gamma \vdash \bot \prop
   \end{prooftree}
\qquad
\begin{prooftree}
 \Gamma \vdash \phi \prop
\quad
\Gamma \vdash \Phi \Rightarrow \bot
\justifies
\Gamma \vdash \Phi \Rightarrow \phi
\end{prooftree} \]

\[ \begin{prooftree}
 \Gamma \vdash \phi \prop
\quad
\Gamma \vdash \psi \prop
\justifies
\Gamma \vdash \phi \supset \psi \prop
\end{prooftree}
\qquad
\begin{prooftree}
 \Gamma \vdash \phi = \phi'
\quad
\Gamma \vdash \psi = \psi'
\justifies
\Gamma \vdash (\phi \supset \psi) = (\phi' \supset \psi')
\end{prooftree} \]

\[ \begin{prooftree}
    \Gamma \vdash \Phi, \phi \Rightarrow \psi
\justifies
\Gamma \vdash \Phi \Rightarrow \phi \supset \psi
   \end{prooftree}
\qquad
\begin{prooftree}
 \Gamma \vdash \Phi \Rightarrow \phi \supset \psi
\quad
\Gamma \vdash \Phi \Rightarrow \phi
\justifies
\Gamma \vdash \Phi \Rightarrow \psi
\end{prooftree} \]

\[ (\mathrm{DN}) \begin{prooftree}
                  \Gamma \vdash \Phi \Rightarrow \neg (\neg \phi)
\justifies
\Gamma \vdash \Phi \Rightarrow \phi
                 \end{prooftree} \]

\[ \begin{prooftree}
\Gamma, x : A \vdash \phi \prop
\justifies
\Gamma \vdash \forall x:A.\phi \prop
   \end{prooftree}
\qquad
\begin{prooftree}
 \Gamma \vdash A = A'
\quad
\Gamma, x : A \vdash \phi = \phi'
\justifies
\Gamma \vdash (\forall x : A.\phi) = (\forall x:A'.\phi')
\end{prooftree} \]

\[ \begin{prooftree}
\begin{array}{c}
    \Gamma \vdash \phi_1 \prop
\quad
\cdots
\quad
\Gamma \vdash \phi_n \prop \\
\Gamma, x : A \vdash \phi_1, \ldots, \phi_n \Rightarrow \psi
\end{array}
\justifies
\Gamma \vdash \phi_1, \ldots, \phi_n \Rightarrow \forall x:A.\psi
   \end{prooftree}
\qquad
\begin{prooftree}
 \Gamma \vdash \Phi \Rightarrow \forall x:A.\phi
\quad
\Gamma \vdash M : A
\justifies
\Gamma \vdash \Phi \Rightarrow [M/x]\phi
\end{prooftree} \]

\subsubsection{The Propositional Universe}
\label{section:propuni}
\[ \begin{prooftree}
    \Gamma \vdash P \p
\justifies
\Gamma \vdash V(P) \prop
   \end{prooftree}
\qquad
\begin{prooftree}
 \Gamma \vdash P = Q
\justifies
\Gamma \vdash V(P) = V(Q)
\end{prooftree} \]

\[ \begin{prooftree}
    \Gamma \vdash \vald
\justifies
\Gamma \vdash \hat{\bot} \p
   \end{prooftree}
\qquad
\begin{prooftree}
 \Gamma \vdash \vald
\justifies
\Gamma \vdash V(\hat{\bot}) = \bot
\end{prooftree} \]

\[ \begin{prooftree}
\Gamma \vdash P \p
\quad
\Gamma \vdash Q \p
\justifies
\Gamma \vdash P \hat{\supset} Q \p
   \end{prooftree}
\qquad
\begin{prooftree}
 \Gamma \vdash P = P'
\quad
\Gamma \vdash Q = Q'
\justifies
\Gamma \vdash (P \hat{\supset} Q) = (P' \hat{\supset} Q')
\end{prooftree} \]

\[ \begin{prooftree}
    \Gamma \vdash P \p
\quad
\Gamma \vdash Q \p
\justifies
\Gamma \vdash V(P \hat{\supset} Q) = (V(P) \supset V(Q))
   \end{prooftree} \]

\[ \begin{prooftree}
    \Gamma, x : T(M) \vdash P \p
\justifies
\Gamma \vdash \hat{\forall} x : M. P \p
   \end{prooftree}
\qquad
\begin{prooftree}
 \Gamma \vdash M = M' : U\
\quad
\Gamma, x : T(M) \vdash P = P'
\justifies
\Gamma \vdash (\hat{\forall} x : M.P) = (\hat{\forall} x : M'.P')
\end{prooftree} \]

\[ \begin{prooftree}
    \Gamma, x : T(M) \vdash P \p
\justifies
\Gamma \vdash V(\hat{\forall} x :M.P) = (\forall x : T(M). V(P))
   \end{prooftree} \]

\subsubsection{Equality}
\label{section:equality}
\[ \begin{prooftree}
    \Gamma \vdash M_1 : T(N)
\quad
\Gamma \vdash M_2 : T(N)
\justifies
\Gamma \vdash (M_1 =_N M_2) \prop
   \end{prooftree}
\qquad
\begin{prooftree}
\begin{array}{c}
 \Gamma \vdash N = N' : U \\
 \Gamma \vdash M_1 = M_1' : T(N)
\\
\Gamma \vdash M_2 = M_2' : T(A)
\end{array}
\justifies
\Gamma \vdash (M_1 =_N M_2) = (M_1' =_N M_2')
\end{prooftree} \]

\[ \begin{prooftree}
\begin{array}{c}
 \Gamma \vdash \phi_1 \prop
\quad
\cdots
\quad
\Gamma \vdash \phi_n \prop \\
\Gamma \vdash M : T(N)
\end{array}
\justifies
\Gamma \vdash \phi_1, \ldots, \phi_n \Rightarrow M =_N M
   \end{prooftree} \]

\[ (\mathrm{subst}) \; \begin{prooftree}
\begin{array}{c}
    \Gamma, x : T(N) \vdash \phi \prop \\
\Gamma \vdash \Phi \Rightarrow M_1 =_N M_2
\quad
\Gamma \vdash \Phi \Rightarrow [M_1/x]\phi
\end{array}
\justifies
\Gamma \vdash \Phi \Rightarrow [M_2/x]\phi
   \end{prooftree} \]

\[ \begin{prooftree}
    \Gamma \vdash M_1 : T(N)
\quad
\Gamma \vdash M_2 : T(N)
\justifies
\Gamma \vdash (M_1 \hat{=}_N M_2) \p
   \end{prooftree}
\qquad
\begin{prooftree}
\begin{array}{c}
 \Gamma \vdash M_1 = M_1' : T(N)
\\
\Gamma \vdash M_2 = M_2' : T(N)
\end{array}
\justifies
\Gamma \vdash (M_1 \hat{=}_N M_2) = (M_1' \hat{=}_N M_2')
\end{prooftree} \]

\[ \begin{prooftree}
    \Gamma \vdash M_1 : T(N)
\quad
\Gamma \vdash M_2 : T(N)
\justifies
\Gamma \vdash V(M_1 \hat{=}_N M_2) = (M_1 =_N M_2)
   \end{prooftree} \]

\subsubsection{Differences from Previous Presentation}

The above presentation differs from the one in \cite{al:wpcm2} in a few respects.  In that paper, we constructed $\LTTW$ within the logical framework LF$'$.
Here, we have presented $\LTTW$ as a separate, stand-alone formal system.  
The constant Peirce in \cite{al:wpcm2} has been replaced with the rule (DN),
the constant $I_\Rightarrow$
has been replaced with the rule ($\etaA$), and the constant $I_\times$
has been replaced with ($\etaX$).

It is not difficult to show that the two presentations are equivalent.
These changes have been made in order to simplify the definition of the interpretations in Section \ref{section:conservativity}.

In \cite{al:wpcm2}, we introduced a proposition `prop', and used the proofs of `prop' as the names of the small propositions.  We also discussed the possibility of making `prop' a type.  In this paper, we have taken a neutral option: we have used a separate judgement form $\Gamma \vdash P \p$.  The system we present here can be embedded in both the system that has `prop' a proposition, and the system that has `prop' a type.  It can be shown that these two embeddings are conservative.

\subsection{$\LTTO$}
\label{app:ltto}
The subsystem $\LTTO$ is formed from $\LTTW$ by making the following changes.
\begin{enumerate}
\item 
Whenever the rules $(\EN)$, $(\EN=)$, $(\EN0)$ or $(\EN\s)$ are used, the type $A$ must have the form $T(K)$.
\item
Whenever the rule $(\IndN)$ is used, the proposition $\phi$ must have the form $V(P)$.
\item
Whenever the rule (subst), $(\etaX)$ or $(\etaA)$ is used, then for every quantifier $\forall x : A$ in the proposition $\phi$, the type $A$ must not contain the symbol $U$.
\item
The following rule of deduction is added:
\[ (P3) \; \begin{prooftree}
    \Gamma \vdash \phi_1 \prop
\quad
\cdots
\quad
\Gamma \vdash \phi_n \prop
\quad
\Gamma \vdash M : \N
\justifies
\Gamma \vdash \phi_1, \ldots, \phi_n \Rightarrow \neg (0 =_{\hat{\N}} \s M)
   \end{prooftree} \]
\end{enumerate}

\subsection{$\LTTOs$}
\label{app:lttos}
We say a proposition $\phi$ is \emph{analytic} iff, for every quantifier $\forall x:A$ in $\phi$, either $A \equiv T(M)$ for some $M$, or $A \equiv \Set{\N}$.

The subsystem $\LTTO$ is formed from $\LTTW$ by making the following changes.
\begin{enumerate}
\item 
Whenever the rules $(\EN)$, $(\EN=)$, $(\EN0)$ or $(\EN\s)$ are used, the type $A$ must have the form $T(K)$.
\item
Whenever the rule $(\IndN)$ is used, the proposition $\phi$ must be analytic.
\item
Whenever the rule (subst), $(\etaX)$ or $(\etaA)$ is used, the proposition $\phi$ must have the form $V(P)$.
\item 
The rule of deduction $(P3)$ is added.
\end{enumerate}

\section{Proof of the Soundness Theorems}
\setcounter{equation}{0}
\renewcommand{\theequation}{B.\arabic{equation}}
We present here the proofs of two of the Soundness Theorems in this paper.

\subsection{Proof of Theorem \ref{thm:soundness}}
\label{app:pfsound}

We begin by proving the following properties of our interpretation:
\begin{lm}
\label{weakening'}
If $\Delta \subseteq \Delta'$, then $\brackets{A}_\Delta \subseteq \brackets{A}_{\Delta'}$ and $(\sim_\Delta^A) \subseteq (\sim_{\Delta'}^A)$.
\end{lm}

\begin{pf}
The proof is by induction on $A$.
\end{pf}

\begin{lm}$ $
\label{lm:properties}
\begin{enumerate}
\item
\label{sub}
Let $M$ be a term and $X$ an expression of $\mathcal{A}_{n+1}$.  Let $v' = v[x:=\round{M}^v]$.
If $\round{M}^v$ is defined, and $\round{X}^{v'}$ is defined, then $\round{[M/x]X}^v$ is defined, and
$ \round{[M/x]X}^v = \round{X}^{v'}$.
\item
Given a term $M$ of $\mathcal{A}_n$ and expression $X$ of $\mathcal{A}_{n+1}$, we have
$[M/x]\round{X}^v \simeq \round{X}^u$
where, for all $y \in \dom v$, $u(y) \equiv [M/x]v(y)$.
\item
\label{substitution}
If $\round{M}^v$ and $\round{X}^{v[x:=x]}$ are defined, then $\round{[M/x]X}^v$ is defined, and
$\round{[M/x]X}^v \equiv [\round{M}^v / x] \round{X}^{v[x:=x]}$
\item
If $v(x) = v'(x)$ for all $x \in \FV{M}$, then
$\round{X}^v = \round{X}^{v'}$.
\item
\label{weakening}
Suppose $(\Delta, \Phi, v) \models \phi$.  If $\Delta \subseteq \Delta'$, $\Phi \subseteq \Phi'$, and $v(x) = v'(x)$ for all $x \in \FV{\phi}$, then $(\Delta', \Phi', v') \models \phi$.
\item
\label{substitution'}
$(\Delta, \Phi, v) \models [M/x]\phi$ iff $(\Delta, \Phi, [M/x]v) \models \phi$.
\end{enumerate}
\end{lm}

\begin{pf}
Part 1 is proven by induction on $X$, and part 2 by induction on $N$.  Part 3 follows simply from the first two.
The remaining parts are proven by induction on $X$ or $\phi$.
\end{pf}

Theorem \ref{thm:soundness} is now proven by induction on derivations.  We deal with five cases here.
\begin{enumerate}
\item
Consider the case of the rule of deduction
\[ \begin{prooftree}
    \Gamma, x : A \vdash M : B
\quad
\Gamma \vdash N : A
\justifies
\Gamma \vdash (\lambda x:A.(M:B))(N)_{A \rightarrow B} = [N/x]M : B
   \end{prooftree} \]
By the induction hypothesis, we have
\[ \Delta, x:A \vdash_n \round{M}^{v[x:=x]} : B, \qquad \Delta \vdash_n \round{N}^v : A \]
and we must show $\Delta \vdash_n \round{(\lambda x:A.M)(N)}^v = \round{[N/x]M}^v : B$.

Suppose $d(A \rightarrow B) \leq n$.  Then we have
\[ \Delta \vdash_n (\lambda x:A.\round{M}^{v[x:=x]})(\round{N}^v) = [\round{N}^v / x]\round{M}^{v[x:=x]} : B \enspace . \]
By the two claims above, we have
$[\round{N}^v / x]\round{M}^{v[x:=x]} \equiv \round{[N/x]M}^v$
and the required judgement follows.

Suppose now $d(A \rightarrow B) = n + 1$.  We must show
$\Delta \vdash \round{(\lambda x:A.M)(N)}^v = \round{[M/x]N}^v : B$
But
\begin{eqnarray*}
 \round{(\lambda x:A.M)(N)}^v & \equiv & \round{\lambda x:A.M}^v @ \round{N}^v \\
& \equiv & \langle x, \round{M}^{v[x:=x]} \rangle @ \round{N}^v \\
& \equiv & [\round{N}^v / x]\round{M}^{v[x:=x]} \\
& \equiv & \round{[M/x]N}^v
\addtocounter{equation}{1} \end{eqnarray*}
and so the required judgement is
\[ \Delta \vdash [\round{N}^v / x]\round{M}^{v[x:=x]} = [\round{N}^v / x]\round{M}^{v[x:=x]} : B \]
which is derivable in $\mathcal{A}_n$.
\item
Consider the rule of deduction
\[ \begin{prooftree}
\Gamma \vdash \Psi \Rightarrow \forall x :A. \psi \qquad
\Gamma \vdash M : A
\justifies
\Gamma \vdash \Psi \Rightarrow [M/x]\psi
   \end{prooftree} \]
Suppose $(\Phi, \Delta, v)$ satisfies each member of $\Psi$.  Then $(\Phi, \Delta, v) \models \forall x:A.\psi$.  We also have $\round{M}^v \in \brackets{A}_\Delta^v$.

If $d(\forall x:A.\psi) \leq n$, then we have $\Delta \vdash \Phi \Rightarrow \forall x : A. \round{\psi}^v$ and $\Delta \vdash \round{M}^v : A$, hence $\Delta \vdash \Phi \Rightarrow [\round{M}^v / x]\round{\psi}^{v[x:=x]}$, and this is the judgement required by Lemma \ref{lm:properties}.\ref{substitution}.

If $d(\forall x:A.\psi) = n+1$, then we have $(\Phi, \Delta, v[x:=\round{M}^v]) \models \psi$.  Hence $(\Phi, \Delta, v) \models [M/x]\psi$ by Lemma \ref{lm:properties}.\ref{substitution'} as required.

\item
Consider the rule of deduction
\[ \begin{prooftree}
\Gamma \vdash \psi \prop
\qquad
\Gamma \vdash \Psi \Rightarrow \bot
\justifies
\Gamma \vdash \Psi \Rightarrow \psi
\end{prooftree} \]
For this case, we need the result:
\begin{quote}
If $\Delta \vdash \Phi \Rightarrow \bot$ then $(\Delta, \Phi, v) \models \psi$ for every proposition $\psi$ of $\mathcal{A}_{n+1}$.
\end{quote}
This is proven by induction on $\psi$.
\item
Consider the rule of deduction
\[ (DN) \; \begin{prooftree}
\Gamma \vdash \Psi \Rightarrow \neg \neg \psi
\justifies
\Gamma \vdash \Psi \Rightarrow \psi
\end{prooftree} \]
For this case, we need the result:
\begin{quote}
If $(\Delta, \Phi, v) \models \neg \neg \phi$ then $(\Delta, \Phi, v) \models \phi$.
\end{quote}
If $d(\phi) \leq n$, we have
\begin{eqnarray*}
\Delta & \vdash & \Phi \Rightarrow \neg \neg \round{\psi}^v \\
\therefore \Delta & \vdash & \Phi \Rightarrow \round{\psi}^v & (DN)
\addtocounter{equation}{1} \end{eqnarray*}

If $d(\phi) = n+1$ and $\phi \equiv \psi \supset \chi$, we have that
\begin{equation}
\label{negneg1}
(\Delta, \Phi, v) \models \neg \neg (\psi \supset \chi) \enspace .
\end{equation}
Suppose $\Delta_1 \supseteq \Delta$, $\Phi_1 \supseteq \Phi$, and 
\begin{equation}
\label{negneg2}
(\Delta_1, \Phi_1, v) \models \psi \enspace .
\end{equation}
We must show $(\Delta_1, \Phi_1, v) \models \chi$.  By the induction hypothesis, it is sufficient to prove $(\Delta_1, \Phi_1, v) \models \neg \neg \chi$.  So suppose $\Delta_2 \supseteq \Delta_1$, $\Phi_2 \supseteq \Phi_1$, and
\begin{equation}
\label{negneg3}
(\Delta_2, \Phi_2, v) \models \neg \chi \enspace .
\end{equation}
We must show $(\Delta_2, \Phi_2, v) \models \bot$.  By (\ref{negneg1}), it is sufficient to prove that $(\Delta_2, \Phi_2, v) \models \neg(\psi \supset \chi)$.  So suppose $\Delta_3 \supseteq \Delta_2$, $\Phi_3 \supseteq \Phi_2$, and
\begin{equation}
\label{negneg4}
(\Delta_3, \Phi_3, v) \models \psi \supset \chi \enspace .
\end{equation}
We have $(\Delta_3, \Phi_3, v) \models \psi$ by Lemma \ref{lm:properties}.\ref{weakening}, so $(\Delta_3, \Phi_3, v) \models \chi$, and hence $(\Delta_3,
\Phi_3, v) \models \bot$ by (\ref{negneg3}), as required.

The case $d(\phi) = n + 1$ and $\phi \equiv \forall x:A.\psi$ is similar.
\item
Consider the case of the rule of deduction ($\IndN$):
\[ \begin{prooftree}
\begin{array}{cc}
\Gamma, x : \N \vdash V(P) \prop & \Gamma \vdash N : \N \\
\Gamma \vdash \Phi \Rightarrow V([0/x]P) & \Gamma, x : \N \vdash \Phi, V(P) \Rightarrow V([\s x / x]P)
\end{array}
\justifies
\Gamma \vdash \Phi \Rightarrow V([N/x]P)
\end{prooftree} \]
This follows by applying ($\IndN$) in $\mathcal{A}_n$.  Note that it is important here that $V(P)$ must be a small proposition.
\end{enumerate}

\subsection{Proof of Theorem \ref{thm:soundness'}}
\label{app:pfsound'}
We begin by proving 
\begin{lm}
If $\Delta \subseteq \Delta'$, then $\brackets{A}_\Delta^v \subseteq \brackets{A}_{\Delta'}^v$ and $(\sim_{\Delta v}^A) \subseteq (\sim_{\Delta' v}^A)$.
\end{lm}

\begin{pf}
Similar to Lemma \ref{weakening'}.
\end{pf}

We prove that Lemma \ref{lm:properties} holds for our new translation.  The proof is similar.

Theorem \ref{thm:soundness'} is now proven by induction on derivations.  We deal with one case here: the rule of deduction
\[ (\EN \s) \; \begin{prooftree}
\begin{array}{cc}
                \Gamma, x : \N \vdash T(K) \type
& \Gamma \vdash L : T([0/x]K) \\
\Gamma, x : \N, y : T(K) \vdash M : T([\s x/x]K)
& \Gamma \vdash N : \N
\end{array}
\justifies
\begin{array}{l}
\Gamma \vdash \EN([x]T(K), L, [x,y]M, \s N) \\ \quad = [N/x, \EN([x]T(K), L, [x,y]M, N)/y]M : T([\s N/x]K)
\end{array}
               \end{prooftree} \]
Let $v$ be a $\Delta$-valuation of $\Gamma$.  Inverting, the derivation includes $\Gamma, x : \N \vdash K : U$, and so the induction hypothesis gives us $\round{K}^{v[x:=J]} \in \mathscr{S}$
whenever $\Delta \vdash J : \N$.  Let us define
\begin{eqnarray*} S(J) & = & \round{K}^{v[x:=J]} \\
 D_J & \equiv & \decode_{S(J)} \\
C_J & \equiv & \code_{S(J)} \\
F(J) & \equiv & \EN([x]T(K), L, [x,y]M, J)
\addtocounter{equation}{1} \end{eqnarray*}
We have the following chain of equalities provable in $\Tw$:
\begin{eqnarray*}
\lefteqn{\round{F(\s N)}^v} \\
 & \equiv & D_{\s \round{N}} (\R (C_0(\round{L}), [x,y]C_{\s x}(\round{M}^{v[x:=x, y:=D_x(y)]}), \s \round{N}^v)) \\
& = & D_{\s \round{N}}(C_{\s \round{N}}(\round{M}^{v[x:=\round{N}, y:=\round{F(N)}]})) \\
& = & \round{M}^{v[x:=\round{N}, y:=\round{F(N)}]} \\
& \equiv & \round{[N/x, F(N)/y]M}^v
\addtocounter{equation}{1} \end{eqnarray*}
as required.
\end{document}